\documentclass[conference]{IEEEtran}

\usepackage{cite}
\usepackage{amsmath,amssymb,amsfonts}
\usepackage{algorithmic}
\usepackage{graphicx}
\usepackage{textcomp}
\usepackage{xcolor}

\usepackage{verbatim}  %

\usepackage{bm} %
\usepackage{booktabs} %
\usepackage{multirow} %
\usepackage{wrapfig} %
\usepackage{subcaption} %
\usepackage[utf8]{inputenc} %

\AtBeginDocument{%
  \providecommand\BibTeX{{%
    \normalfont B\kern-0.5em{\scshape i\kern-0.25em b}\kern-0.8em\TeX}}}

\begin{document}

\title{Reliable Geofence Activation with Sparse and Sporadic Location Measurements: Extended Version}

\author{\IEEEauthorblockN{Kien Nguyen}
\IEEEauthorblockA{\textit{Department of Computer Science} \\
\textit{University of Southern California}\\
Los Angeles, CA, USA \\
kien.nguyen@usc.edu}
\and
\IEEEauthorblockN{John Krumm}
\IEEEauthorblockA{\textit{Microsoft Research} \\
\textit{Microsoft Corporation}\\
Redmond, WA, USA \\
jckrumm@microsoft.com}
} 

\maketitle

\begin{abstract}
Geofences are a fundamental tool of location-based services. A geofence is usually activated by detecting a location measurement inside the geofence region. However, location measurements such as GPS often appear sporadically on smartphones, partly due to weak signal, or privacy preservation, because users may restrict location sensing, or energy conservation, because sensing locations can consume a significant amount of energy. These unpredictable, and sometimes long, gaps between measurements mean that entry into a geofence can go completely undetected. In this paper we argue that short term location prediction can help alleviate this problem by computing the probability of entering a geofence in the future. Complicating this prediction approach is the fact that another location measurement could appear at any time, making the prediction redundant and wasteful. Therefore, we develop a framework that accounts for uncertain location predictions and the possibility of new measurements to trigger geofence activations. Our framework optimizes over the benefits and costs of correct and incorrect geofence activations, leading to an algorithm that reacts intelligently to the uncertainties of future movements and measurements.\footnote{A short version of this paper appears in the proceedings of the 23rd IEEE International Conference on Mobile Data Management MDM 2022.} 
\end{abstract}


\section{Introduction}
Geofences are virtual geographic regions used to trigger certain actions upon entry or exit. A typical example is a region near a store where an advertiser may want to deliver ads to the phones of people in the region. When someone with a location-sensitive device enters a geofence, some action is automatically triggered.

\begin{figure}[htbp!]
\centering
\includegraphics[width=0.95\linewidth]{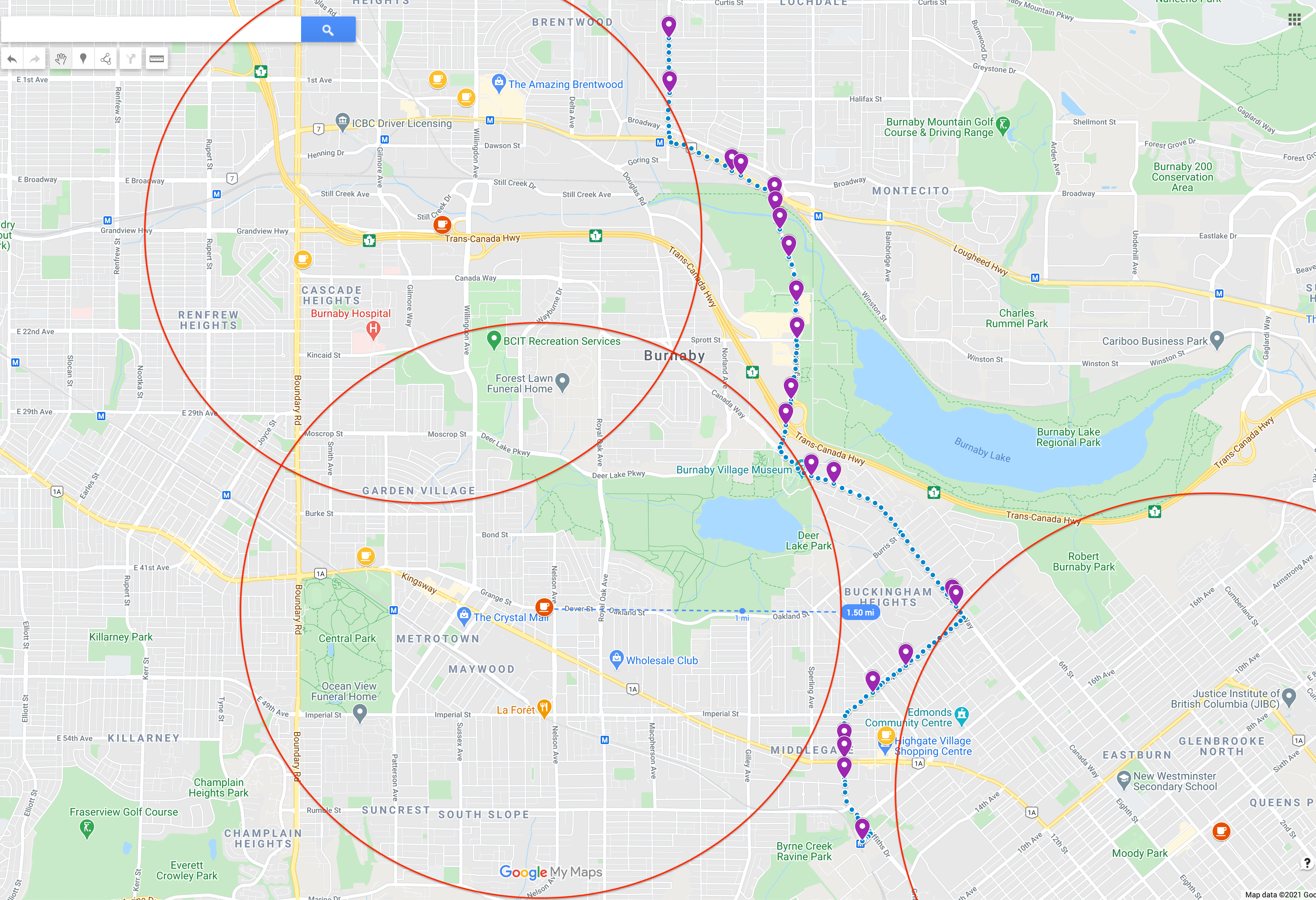}
\caption{A trajectory passing through three goefences can be missed when relying on only discrete location samples.\label{fig:geofence-problem}}
\end{figure}

One problem with geofences is they may fail to activate if there is no location measurement taken inside the geofence, even if the person passes through. For example, in Figure~\ref{fig:geofence-problem}, a full trajectory of a person is shown in blue dots where each blue dot is five seconds apart from the ones before and after. The trajectory passes through three geofences with a 1.5-mile radius from three different Starbucks stores. However, when the trajectory is subsampled to have, in expectation, one measurement every minute to simulate the case where measurements come sporadically, none of the remaining measurements shown with purple pins was found within the three geofences (The data and subsampling process are described in Section~\ref{sec:experimental-data}). Thus, relying solely on the existence of measurements inside a geofence would fail to activate in this case, although the person passed through all three.

\begin{figure}[htbp!]
\centering
\includegraphics[width=.9\linewidth]{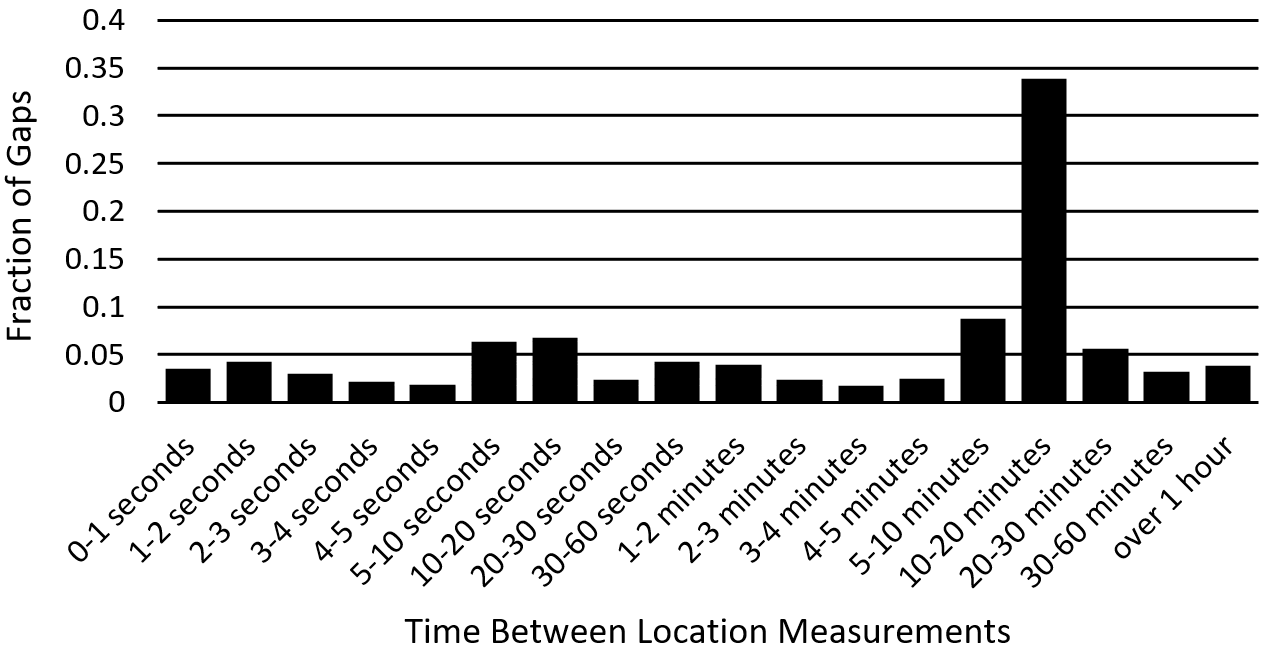}
\caption{The time between location measurements can be long enough to miss a geofence.\label{fig:time-between-location-measurements-1000-random-Safegraph-1-June-2019}}
\end{figure}

This problem can occur whenever the gaps between measurements are large enough to miss a geofence, including when a person is walking or in a vehicle. With a high enough sampling rate, this is not a problem. However, geofence applications may not able to proactively trigger new measurements. One reason is when the location signal is too weak, e.g., inside a tunnel or surrounded by high buildings. Another reason is that user may only allow such applications to monitor location readings after a time delay due to, e.g., privacy concerns. Yet another reason is that sensing location, especially with GPS, drains a phone's battery. For instance, Liu~\emph{et al.}~\cite{liu2012energy} estimate that running a GPS receiver continuously will drain a phone's battery in about six hours. Therefore, some power management programs of the device may restrict, e.g., the frequency of location sensing. This leads to conservative sensing, where measurements are often relatively far apart in time. As an example, Figure~\ref{fig:time-between-location-measurements-1000-random-Safegraph-1-June-2019} shows a  histogram of time spans between measurements for a random sample of 1000 users (described in detail later in Section~\ref{sec:experimental-data}). It shows that
over half the points are separated by five minutes or more, making it easy to miss a geofence depending on the geofence's size and the user's speed. Our results in Section~\ref{sec:experiments} show that the current practice of waiting for a measurement to trigger a geofence performs poorly.

Short term location prediction can help alleviate this problem by giving the probability that the person will enter the geofence in the near future. However, there are two issues: (1) probability by itself may not be sufficient to make the optimal, benefit-maximizing decision of activating the geofence or not, and (2) a new measurement could arrive at any time, wasting prediction computations that go beyond that time.

In this paper we develop a novel framework that uses short term location prediction to solve the problem of triggering geofences with sparse and sporadic location measurements. Our framework takes into account both the predictive uncertainty for decision making and the potentially wasted computation issues. 
The fundamental approach is to use decision theory as a principled way to manage the trade-off between costs and benefits of acting or waiting. Decision theory is enabled by computing the probability that a user will intersect the geofence before the next measurement is available using a probabilistic location prediction method.
This is a new approach for deciding whether or not to trigger a geofence, reflecting richer, more subtle reasoning than the traditional method of passively waiting for a point to appear inside the geofence. In addition, when we use location prediction, decision theory helps bridge the gap between the accuracy of location prediction and the benefit/penalty of geofence triggers in a principled way.
We further reason explicitly about the temporally sporadic nature of location measurements by modeling their arrival times as Poisson distributed. This reasoning gives us a principled way to stop the prediction process, thus avoiding redundant predictions. We show how our approach is superior to the baseline technique of using only the given measurements without considering the possibility of missing a geofence. To the best of our knowledge, in the context of geofences, neither probabilistic location prediction, nor decision theory, nor reasoning about sporadic measurements has appeared in the research literature before.

For both privacy and energy conservation, our approach is designed to be simple enough to run on the user's local device rather than transmitting any location data. Likewise, it does not trigger any new measurements, relying instead on opportunistic measurements that are made available by other processes on the device.

Specifically, our new research contributions are:
\begin{itemize}
   \item Probabilistically predicting geofence intersections
   \item Modeling costs/benefits of geofence activations with decision theory
   \item Reasoning about temporally sporadic location measurements with a Poisson distribution
   \item Extensive experiments over different settings for location measurements and algorithmic parameters
\end{itemize}

\section{Related Work}
While it seems obvious to use location prediction with geofences, there is surprisingly little research on the topic. We are solving the problem of missing a user's entrance into a geofence due to low sampling rates.
In~\cite{zimbelman2017hazards}, Zimbelman~\emph{et al.} characterize a similar problem of trying to detect when a moving geofence (\emph{e.g.} around a person) intersects a static location. Their experiments looked at the effects of geofence speed, geofence size, location sampling interval, and intersection angle on the delay of detecting geofence intersections. Fattepur~\emph{et al.}~\cite{fattepur2016solution} present a state transition algorithm for a GNSS chipset that does simple reasoning about geofences, but does not use location prediction. The work most closely related to ours is from Nakagawa~\emph{et al.}~\cite{nakagawa2013variable}. Their solution to the problem of missing geofences is to adaptively increase the location sampling rate when the user is getting closer to a geofence. The estimated distance to the geofence is based on a simple, deterministic prediction from the last two measured locations. In contrast, our method uses a probabilistic prediction, decision theory, and does not trigger new measurements.

While there has not been much research on geofence intersection detection, there is a large literature on location prediction, which is one of the central components of our approach. A survey of location prediction approaches appears in~\cite{cheng2003location}. We are particularly interested in short-term location prediction that gives future location estimates in the relatively short span between measurements. For instance, the median time between measurements for the 1000 random users represented in Figure~\ref{fig:time-between-location-measurements-1000-random-Safegraph-1-June-2019} was about 8.5 minutes. Our framework requires a probabilistic location prediction in order to accommodate decision theory about whether or not to activate a geofence. In the realm of short term, probabilistic location prediction, a classic example is the Kalman filter~\cite{kalman1960new}, which creates a Gaussian-distributed prediction as part of its measurement update algorithm. The particle filter~\cite{doucet2001introduction} also has a prediction step, as does the unscented Kalman filter~\cite{wan2000unscented}.

Location prediction is one component of our approach, but our main contribution is a framework for optimally reasoning about when to trigger a geofence in light of the inherent uncertainty about the location prediction and uncertainty about when the next measurement will be available. We handle the uncertainty using decision theory, which lets us reason not about accuracy, but about the ultimate costs and benefits of decisions which are likely more relevant and explainable to a geofence owner.
\section{Decision Theory with Location Predictions} \label{sec:theory}
This section introduces our geofence decision theory inspired by the uncertainty of users' future locations, our use and modification of Gaussian process for probabilistic location prediction, and our reasoning for the temporally sporadic nature of location measurements.

\subsection{Decision Theory} \label{subsec:decision_theory}
\begin{figure}[ht!]
\centering
\includegraphics[width=.4\linewidth]{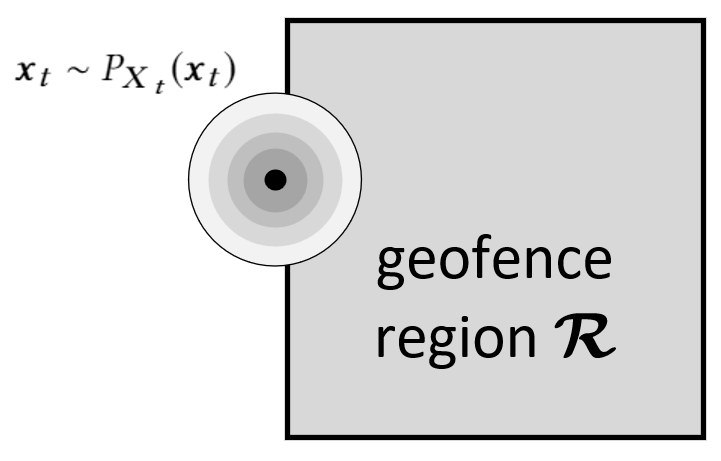}
\caption{With an uncertain location $\boldsymbol{x}(t) \sim P_{\boldsymbol{X}(t)}\big(\boldsymbol{x}(t)\big)$, there is a probability that the user is inside $\mathcal{R}$.\label{fig:geofence-probability}}
\end{figure}

A geofence triggers some action when a user enters\footnote{An action may be triggered when a user \emph{leaves}. For that, we imagine a complementary "entry" geofence covering everywhere except the region of the "exit" geofence.}. However, a user's location typically has some uncertainty due to measurement noise or prediction error. This leads to uncertainty about whether or not the user is inside the geofence. We represent a user's two-dimensional location coordinates at time $t$ as the vector $\boldsymbol{x}(t) = [x(t), y(t)]^T$, distributed according to the probability distribution $P_{\boldsymbol{X}(t)}\big(\boldsymbol{x}(t)\big)$. If the geofence region is represented by $\mathcal{R}$, the scalar probability that the user is inside the geofence is 
\begin{equation} \label{eq:probability_inside_geofence}
p_\mathcal{R}(t)=\int_\mathcal{R} P_{\boldsymbol{X}(t)}\big(\boldsymbol{x}(t)\big) d\boldsymbol{x}
\end{equation}
\noindent as illustrated in Figure~\ref{fig:geofence-probability}.

Based on $p_\mathcal{R}(t)$, the geofence can be programmed to either act or wait: acting means the geofence triggers some action, e.g. the delivery of an advertisement to the user, while waiting means nothing is triggered. A payoff matrix captures the value $V(t)$ of acting or waiting depending on whether or not the user is inside the geofence, shown in Table~\ref{tab:payoff-matrix}. 
The value of acting when the user is inside $\mathcal{R}$ is $\beta$, which would normally be positive, reflecting the intended functioning of the geofence. The value of the two error conditions are acting when the user is outside ($\delta$) and waiting when the user is inside ($\alpha$). Both of these values would likely be negative. Waiting while the user is outside is the correct decision, but the payoff in this case would normally be zero. The exact values of the elements of the payoff matrix depend on the scenario. For advertising, the values would depend on the cost of delivering an ad and the response rate. These values are normally proprietary and beyond the scope of this paper, although this would be an interesting extension of our work. We explore different payoff matrix settings in our experiments. Using a payoff matrix instead of raw prediction accuracy has two advantages. First, it allows us to express and evaluate the true costs of mistakes and successes. Second, by using costs, it lets the algorithm optimize for cost rather than raw accuracy, which can lead to different decisions.

\begin{table}[hb!]
\centering
\begin{tabular}{llccl}
\cline{3-4}
                                                & \multicolumn{1}{l|}{}     & \multicolumn{2}{c|}{\textbf{user state}}                                                         &  \\ \cline{3-4}
                                                & \multicolumn{1}{l|}{}     & \multicolumn{1}{c|}{in}                    & \multicolumn{1}{c|}{out}                   &  \\ \cline{1-4}
\multicolumn{1}{|c|}{\multirow{2}{*}{\textbf{decision}}} & \multicolumn{1}{c|}{wait} & \multicolumn{1}{c|}{$\alpha$} & \multicolumn{1}{c|}{0}                     &  \\ \cline{2-4}
\multicolumn{1}{|c|}{}                          & \multicolumn{1}{c|}{act}  & \multicolumn{1}{c|}{$\beta$}  & \multicolumn{1}{c|}{$\delta$} &  \\ \cline{1-4}
                                                &                           & \multicolumn{1}{l}{}                       & \multicolumn{1}{l}{}                       &  \\
                                                &                           & \multicolumn{1}{l}{}                       & \multicolumn{1}{l}{}                       & 
\end{tabular}
\caption{The payoff matrix gives values of decisions depending on the state of the user.}
\label{tab:payoff-matrix}
\end{table}

The expected payoff values, given an activation decision, can be computed from $p_\mathcal{R}(t)$ and the payoff matrix:

\begin{align}
    \mathbb{E}\big[V(t) \mid \textrm{wait}\big]&=\alpha p_\mathcal{R}(t) + 0(1-p_\mathcal{R}(t)) \\ 
        \mathbb{E}\big[V(t) \mid \textrm{act}\big]&=\beta p_\mathcal{R}(t) + \delta (1-p_\mathcal{R}(t))  \\
        \mathbb{E}\big[V(t)\big]&=\textrm{max}\Big(\mathbb{E}\big[V(t) \mid \textrm{wait}\big],\mathbb{E}\big[V(t) \mid \textrm{act}\big]\Big)
    \label{eq:expected_payoff}
\end{align}

\noindent The decision to wait or act corresponds to which has the larger expected value. This changes with time depending on $p_\mathcal{R}(t)$. In our scenario, after the first "act" decision, the geofence is deactivated, disallowing any acts for that user for some time.

This rule for waiting or acting, based on Equation~\ref{eq:expected_payoff}, is a principled way to account for the costs and benefits of acting under uncertainty for geofences, and is our main contribution to the problem of geofence activation.
The values from the payoff matrix allow the geofence owner (\emph{e.g.} advertiser) to quantify the urgency of delivering a message to someone who should receive it ($\beta$), versus the cost of not delivering it to someone who should receive it ($\alpha$), versus the cost of mistakenly delivering it to someone who should not receive it ($\delta$). The probability $p_\mathcal{R}(t)$ accounts for measurement uncertainty, and would be especially applicable for less precise location sensing modalities, e.g. cell towers or WiFi. 
However, we are interested in further dealing with the problem of low sample rate data, when entering a geofence can be completely missed. For this we can use location prediction, described next.

\subsection{Location Prediction} \label{subsec:location_prediction}

\begin{figure}[ht!]
\centering
\includegraphics[width=.9\linewidth]{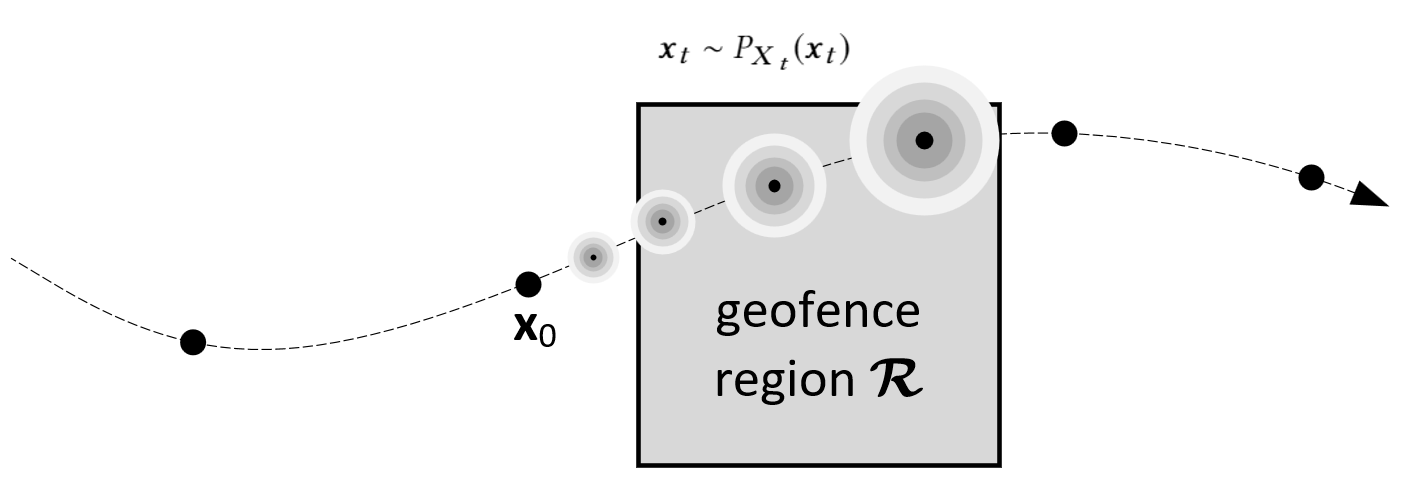}
\caption{$t=0$ is the time of the most recent measurement. Our algorithm makes probabilistic predictions of location to infer the future probability of being inside the geofence.\label{fig:geofence-prediction}}
\end{figure}

We can predict whether or not a user will be inside a geofence by predicting the user's location $\boldsymbol{x}(t)$. Referring to Figure~\ref{fig:geofence-prediction}, we reset the clock to $t=0$ at the most recent location measurement $\boldsymbol{x}_0$, thus $\boldsymbol{x}(0) = \boldsymbol{x}_0$. Then we use an algorithm to create probabilistic predictions $p_\mathcal{R}(t)$ for $t>0$. At each $t>0$, we evaluate Equation~\ref{eq:expected_payoff}, and we "act" the first time $\mathbb{E}\big[V(t) \mid \textrm{act}\big] > \mathbb{E}\big[V(t) \mid \textrm{wait}\big]$. Once we have acted on the user (\emph{e.g.} sent a message) for this geofence, we will not act again for that user on that geofence, or we may not act again until some preset time elapses, such as a few hours.

Our framework accepts any type of probabilistic location prediction. That is, the prediction must produce a distribution $\boldsymbol{x}(t) \sim P_{\boldsymbol{X}(t)}\big(\boldsymbol{x}(t)\big)$ for $t>0$. However, as we aim for mobile applications, we also prefer a light-weight model that can run on-device, even without sending measurements out of the device. Therefore, for our experiments, we predicted location with Gaussian processes (GPs)~\cite{williams1996gaussian}. For a scalar function $x(t)$, a GP implies that any subset of points sampled from the function is distributed according to a multidimensional Gaussian. We create independent GPs for $x(t)$ and $y(t)$ to predict two-dimensional location $\boldsymbol{x}(t) = [x(t),y(t)]^T$.

Several choices are required for implementing a GP. One of them is the standard deviation of the assumed Gaussian noise of the measurements. Gaussian noise is an acceptable approximation of GPS noise~\cite{diggelen2007system}, and we nominally assume a standard deviation of $\sigma_m = 3$~meters. A GP also depends on a scalar covariance function $k(t, t')$ defining how much a measurement $x(t)$ correlates with a measurement $x(t')$. In general, the correlation decreases to zero as $|t-t'|$ gets larger. A common kernel is the squared exponential

\begin{equation}
    k(t, t') = \sigma^2_f \exp \Bigg[-\frac{(t-t')^2}{2l^2}\Bigg]
\end{equation}

\noindent for some choice of parameters $\sigma_f$ and $l$ trained on previous data whenever a new measurement arrives. So, another choice is the number of points prior to $\boldsymbol{x}_0$ used for this training.

Finally, a GP also depends on a mean function $m(t)$ which defines the expected mean values of the data points. Some GPs are assumed to have a zero mean, and some are not. We explore the effects of all these choices in our experiments in Section~\ref{sec:experiments}.

Although a GP can capture the trends of the user's movement to predict future locations, the movement in the short-term may not follow such trends. For example, a right turn made a minute ago might not indicate another right turn. Therefore, we propose an adaptation of the GP for short-term location prediction.
With the intuition that the short-term movement would be affected more by recent movements than by older movements, we aim to make the prediction rely more on the linear extrapolation of the most recent measurements. 
This can be captured by utilizing the mean function. In our adaption, the mean function $m(t)$ of a GP is set to be the line going through the two latest points $x(t_{-1})$ and $x(t_0)$. We denote this as \textit{GP + mean func}. 

Nominally the geofence processor runs the prediction of $\boldsymbol{x}(t)$ forward in time until the value of acting exceeds the value of waiting. Specifically, it looks for the smallest time $\hat{t} > 0$ when $\mathbb{E}\big[V(\hat{t}) \mid \textrm{act}\big] > \mathbb{E}\big[V(\hat{t}) \mid \textrm{wait}\big]$.

We emphasize that although the application of GPs to geofences is novel, location prediction is not the focus of this paper, and other types of probabilistic prediction could be used as well. GPs are particularly well-suited for our application because they give time-dependent, probabilistic predictions which are suitable for our decision theory approach. In the end, however, we evaluate our methods by how much value they return based on the payoff matrices, not raw prediction accuracy. This would make it easier for the adminsitrator of the geofences (\emph{e.g.} advertiser) to understand the ultimate advantages of using our proposed technique.

\subsection{Sporadic Location Measurements}\label{subsec:sporadic_location_measurements}
In Section~\ref{subsec:location_prediction}, we showed how to compute $\hat{t}$, which is the first time when the expected value of acting is predicted to exceed the expected value of waiting. We assume the underlying architecture includes a real time process scheduler. When a new measurement arrives, our algorithm computes $\hat{t}$ and instructs the process scheduler to schedule an "act" for the user/geofence combination at $\hat{t}$.

However, $\hat{t}$ may be very large, especially if the geofence is a long distance from the user, leading to long, possibly infinite, computations of the location predictions for those geofences. In these cases, a new measurement could occur before the computed $\hat{t}$, causing much of the previous computation to become wasted. This section describes a principled way to suspend the prediction.

Assuming the timing of location measurements follows a Poisson process, we can compute the probability of receiving a new measurement as a function of time. A Poisson process is characterized by $\lambda$, which is the average number of events occurring in some predefined interval $\Delta T$. For us, these events are location measurements, and the predefined interval is set arbitrarily to one minute. The parameter $\lambda$ is called the event rate or the rate parameter. The probability of receiving $n$ measurements in the next $t$ minutes is

\begin{equation} \label{eq:poisson_distribution}
P\big(N(t)=n\big) = \frac{(\lambda t)^n e^{-\lambda t}}{n!}
\end{equation}

The probability of having no measurement by time $t$ is $P\big(N(t) = 0\big) = e^{-\lambda t}$, and of receiving at least one measurement by time $t$ is $P\big(N(t) > 0\big) = 1-e^{-\lambda t}$. Our algorithm stops making predictions when the probability of having at least one measurement is sufficiently high, \emph{i.e.} when $P\big(N(t) > 0\big) > 1-\epsilon$, for some small value of $\epsilon$, $0 < \epsilon < 1$. This occurs at time $t^* = -\ln(\epsilon)/\lambda$. 

Whenever a new measurement arrives, our algorithm sets $t=0$. It then computes $t^*$ as described above, which is the cutoff time for making location predictions. Starting at $t=0$, it samples forward in time until it finds a time when the expected value of acting exceeds the expected value of waiting (\emph{i.e.} $\hat{t}$) or until $t$ exceeds $t^*$. If this second condition happens, the algorithm will not schedule the geofence to act, essentially making $\hat{t} = \infty$ and temporarily deactivating the geofence for that user. This process is repeated every time a new measurement arrives, which means the geofence could be reactivated depending on new location data. We believe this approach can also be a principled way to stop the location prediction process in other mobile-related problems such as estimated time of arrival.

\section{Experimental Data} \label{sec:experimental-data}
We tested our algorithm on 530 location trajectories from 165 subjects. Testing on up to 13,824 geofences simultaneously, this means our test data covered over 7 million trajectory-geofence scenarios.

\subsection{Data Source}
The location data came from a commercial aggregator~\footnote{https://www.safegraph.com/} that ingests, cleans, and sells location data gathered from mobile phones. The data comes from individuals using their phone normally, occasionally running applications that trigger location measurements, e.g. weather, web browsing, or navigation applications. This data simulates a geofencing application that is not actively taking location measurements, instead relying passively on measurements triggered by other applications. 

For each of 1000 randomly selected users with at least one data point, we extracted all their data for the date of 1 June 2019 to understand representative statistics on the quantity and frequency of location data that is normally available from a user. As we show below, the data from these 1000 users varied in terms of the number of data points. 

For our experiments, we used the same data source from the same day, but this time we extracted data from the 1000 users with the \emph{most} data points for that day. We refer to these two data sets as the "random" and "high density" sets. The random data set is used to understand representative statistics of available data, while the high density data is used to simulate various data densities to understand how our algorithm performs. Thus, even though the test trajectories originated from high density trips, our controlled subsampling simulated lower sampling rates for our experiments.

\subsection{Data Statistics}
\begin{figure}[ht!]
\centering
\includegraphics[width=.9\linewidth]{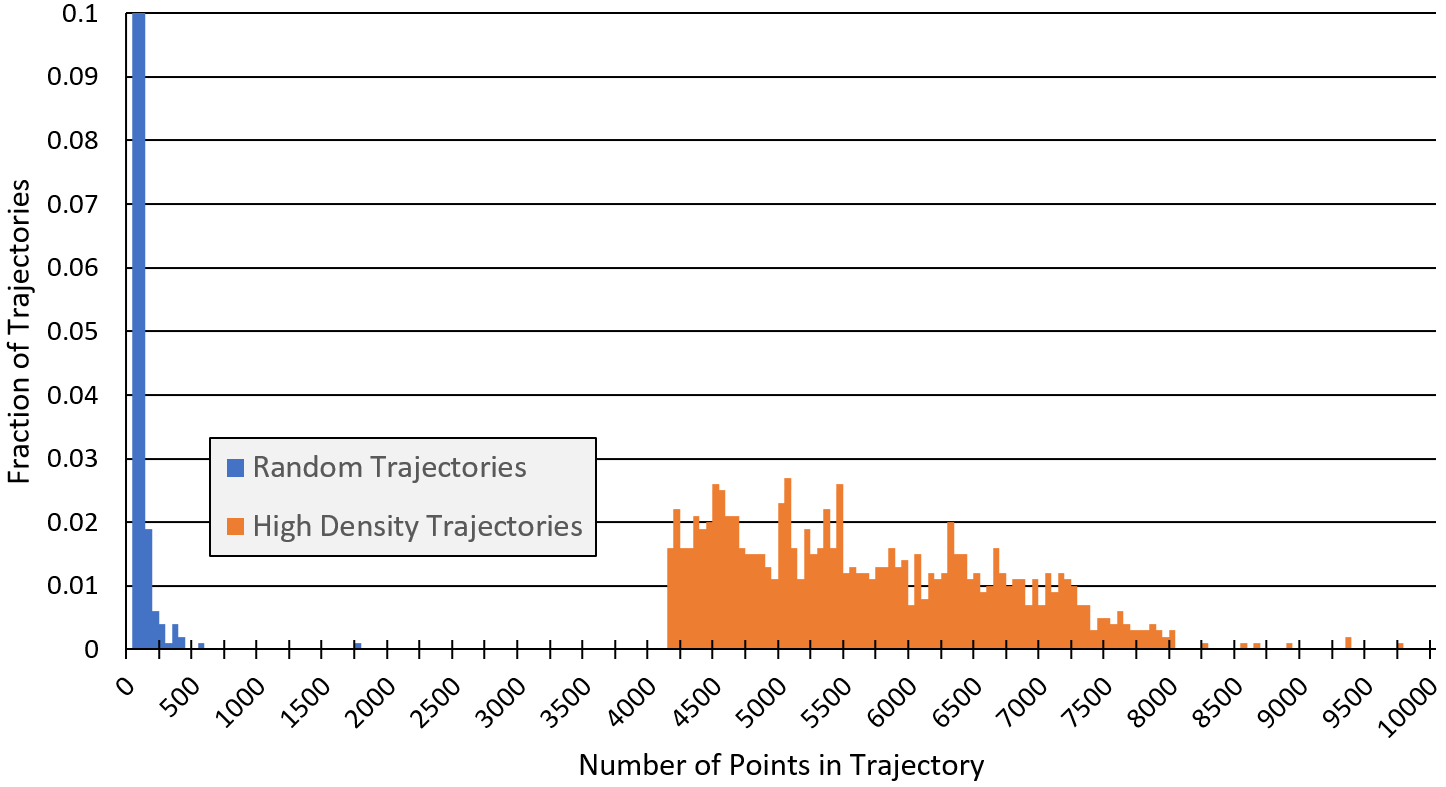}
\caption{Histograms of the number of points in the random and high density trajectories.\label{fig:number-of-points-histogram}}
\end{figure}

Histograms of the number of points in the random and high density sets of trajectories are shown in Figure~\ref{fig:number-of-points-histogram}.
A trajectory $\mathcal{S}$ of a user is a sequence of location measurements $\{\boldsymbol{x}(t_1), \boldsymbol{x}(t_2), \dots, \boldsymbol{x}(t_{N_{\mathcal{S}}})\}$ where $N_{\mathcal{S}}$ is the number of measurements and $\forall i < N_{\mathcal{S}}: t_i < t_{i+1}$.
High density means that, on average, the time gap $|t_i - t_{i+1}|$ is small.
The larger number of points in the high density trajectories enable us to delete some of the points for experiments.

In Section~\ref{subsec:sporadic_location_measurements} we show how to compute a cutoff time for predictions assuming that timing of new location measurements is governed by a Poisson process. The sole parameter of the Poisson distribution (Equation~\ref{eq:poisson_distribution}) is $\lambda$, which is the mean number of events over some unit time. For our experiments, our unit time is one minute. For each of the 1000 random trajectories, we computed the maximum likelihood value of $\lambda$.
The mode of this distribution of $\lambda$'s occurs at 0.04, which corresponds to one measurement every 25 minutes.

We subjected each random trajectory to a statistical test to determine if the timing of the measurements was Poisson distributed. Using the chi-square test~\cite{Zar99} for Poisson distributions, we found that 88.3\% of the random users passed at the $p = 0.05$ level. This indicates that the Poisson distribution is appropriate for modeling the arrival time of location measurements in our data. For the high density data, none of the users passed the statistical test, but these trajectories are outliers, chosen for experimental advantages. Specifically, we downsampled the high density data to simulate different values of $\lambda$ in the Poisson process, described next.

\subsection{Data Pre-Processing} \label{subsec:data-pre-processing}
We processed the high density data to simulate the Poisson processes of the randomly chosen users. Our goal was to find, in the high density data, long sequences of measurements with temporally uniform sampling, since uniform sampling is convenient for down-sampling to measurement times that are Poisson distributed.

For each user, we found the most frequently occurring time gap $\tau$ between temporally adjacent measurements. For the high density data, $\tau$ varied between five to ten seconds. We then split the time series wherever the time gap was not $\tau$, resulting in separate trajectories whose time gaps were uniformly $\tau$.

We then removed trajectories that were short in either time or space. For the time aspect, we aim to have at least five minutes of data for training the GP and five minutes of data for testing. Thus, all trajectories that were shorter than ten minutes were removed. For the space aspect, we removed trajectories where the user was not moving much. We assumed a minimum driving speed of 5m/s (or 18km/h), and removed all trajectories where the distance between the first and last measurements is less than the distance one can travel with minimum driving speed over that time, resulting in total 530 trajectories from 165 unique users.

For these trajectories, we used the first five minutes for training the parameters of the GP. We used the remaining data for testing. Note that the GP does not depend on a large corpus of training data. It computes its own parameters for each trajectory prediction based on the last few points of the measured trajectory.

Then, in the test data period (\emph{i.e.} after five minutes) for each of these trajectories, we computed the first time it entered the geofence, $t_{in}$, and the first time it exited the geofence, $t_{out}$. %
Normally this required interpolation, where we assumed constant, straight line speed between temporally adjacent points.

A subtended degree in latitude and longitude represent different arc lengths on the earth. Therefore, we converted the latitude/longitude coordinates in each trajectory to local Euclidean coordinates $(x, y)$ in meters. For each trajectory, the reference origin  $(lat_0,long_0)$ was arbitrarily chosen as the last training point.

To simulate a Poisson process from our uniformly sampled trajectories, we sample points from the uniform trajectories with a Bernoulli process~\cite{Bonakdarpour01}, which is the discrete-time version of a Poisson process. That is, for every point in the uniform data, we retain it as a measurement with probability $p$ and delete it with probability $1-p$. For the Poisson process, the expected number of events in time $\Delta T$ is $\lambda$. The time between points in the uniform trajectory is $\tau$, meaning there will be $\frac{\Delta T}{\tau}$ points in time $\Delta T$. If we sample with probability $p$, then the expected number of sampled points is $\mathbb{E}[N] = p \frac{\Delta T}{\tau}$ in $\Delta T$. Setting $\mathbb{E}[N] = \lambda$ gives the Bernoulli probability of $p=\lambda \frac{\tau}{\Delta T}$. 

For each trajectory, the training and testing data are sub-sampled separately. We also keep the first and last data points of each part to aid evaluation: the first training data point is kept to make sure that with each configuration for training data for the GPs, we have enough data to provide (\emph{e.g.}, data in the last 5 minutes); the last training data point is kept to be used as the reference origin $(lat_0,long_0)$; and the first and last test data points are kept as the pivot for evaluation, as later shown in Section~\ref{subsec:eval_metric}.

\section{Experiments} \label{sec:experiments}
In this section, we describe our evaluation metric, the baselines, our experimental setup, and our results with some discussion.

\subsection{Evaluation metric}
\label{subsec:eval_metric}
We first describe our evaluation metric to evaluate the effectiveness of our technique for the geofence decision problem. The reason we need a new metric instead of using the whole dense trajectory is because we need a fair way to evaluate proposed methods. Simply comparing a case when the whole dense trajectory is available with when only sporadic data is available (which is our problem setting) is not fair, because our approach is designed to work when data is unavoidably sporadic. Therefore, we propose a "realized value" evaluation score that is analogous to the expected value in Equation~\ref{eq:expected_payoff}. The realized value is the actual payoff, as per the payoff matrix, of using our algorithm to make the act or wait decisions for geofences. Starting first with a single point below, we show how we aggregate the realized value over multiple points in a trajectory $\mathcal{S}$, multiple geofences on the map $\mathcal{R}$, and multiple trajectories in a test set.

We first calculate the realized value, called $V_{\mathcal{R},\mathcal{S}}(t_i)$, for each measurement $\bm{x}(t_i) \in \mathcal{S} \setminus \bm{x}(t_{N_{\mathcal{S}}})$. The last measurement $\bm{x}(t_{N_{\mathcal{S}}})$ is used as a pivot and is not evaluated.

Given $t_{in}$ and $t_{out}$ as the first timestamps that the user enters and exits a geofence $\mathcal{R}$, $t_i$ and $t_{i+1}$ as the timestamps of $i$-th and $(i+1)$-th measurements, and $\hat{t}$ as the predicted timestamp that the geofence should "act" when we receive $\bm{x}(t_i)$, the realized value of the $i$-th measurement $V_{\mathcal{R},\mathcal{S}}(t_i)$ is defined as:
\begin{equation} \label{eq:realized_value_as_a_function_of_t}
    V_{\mathcal{R},\mathcal{S}}(t_i) =
    \begin{cases}
    \alpha, & t_{i+1} < \hat{t} \wedge t_i \leq t_{out} \leq t_{i+1},\\
    \beta, & \hat{t} \leq t_{i+1} \wedge \hat{t} \in [t_{in}, t_{out}],\\
    \delta, & \hat{t} \leq t_{i+1} \wedge \hat{t} \notin [t_{in}, t_{out}].
    \end{cases}
\end{equation}

\noindent where $\alpha$, $\beta$, and $\delta$ come from the payoff matrix in Table~\ref{tab:payoff-matrix}.

Elaborating on the compact notation in Equation~\ref{eq:realized_value_as_a_function_of_t},
the condition $\hat{t} \leq t_{i+1}$ indicates that we "act" before the next measurement arrives. Then, based on the actual user state at the time of acting, we receive reward $\beta$ or penalty $\delta$ if we act when the user is inside (\emph{i.e.}, $\hat{t} \in [t_{in}, t_{out}]$) or outside (\emph{i.e.}, $\hat{t} \notin [t_{in}, t_{out}]$), respectively. This explains the $\beta$ and $\delta$ payoffs in Equation~\ref{eq:realized_value_as_a_function_of_t}.

\begin{figure}[htbp!]
\centering
\includegraphics[width=0.9\linewidth]{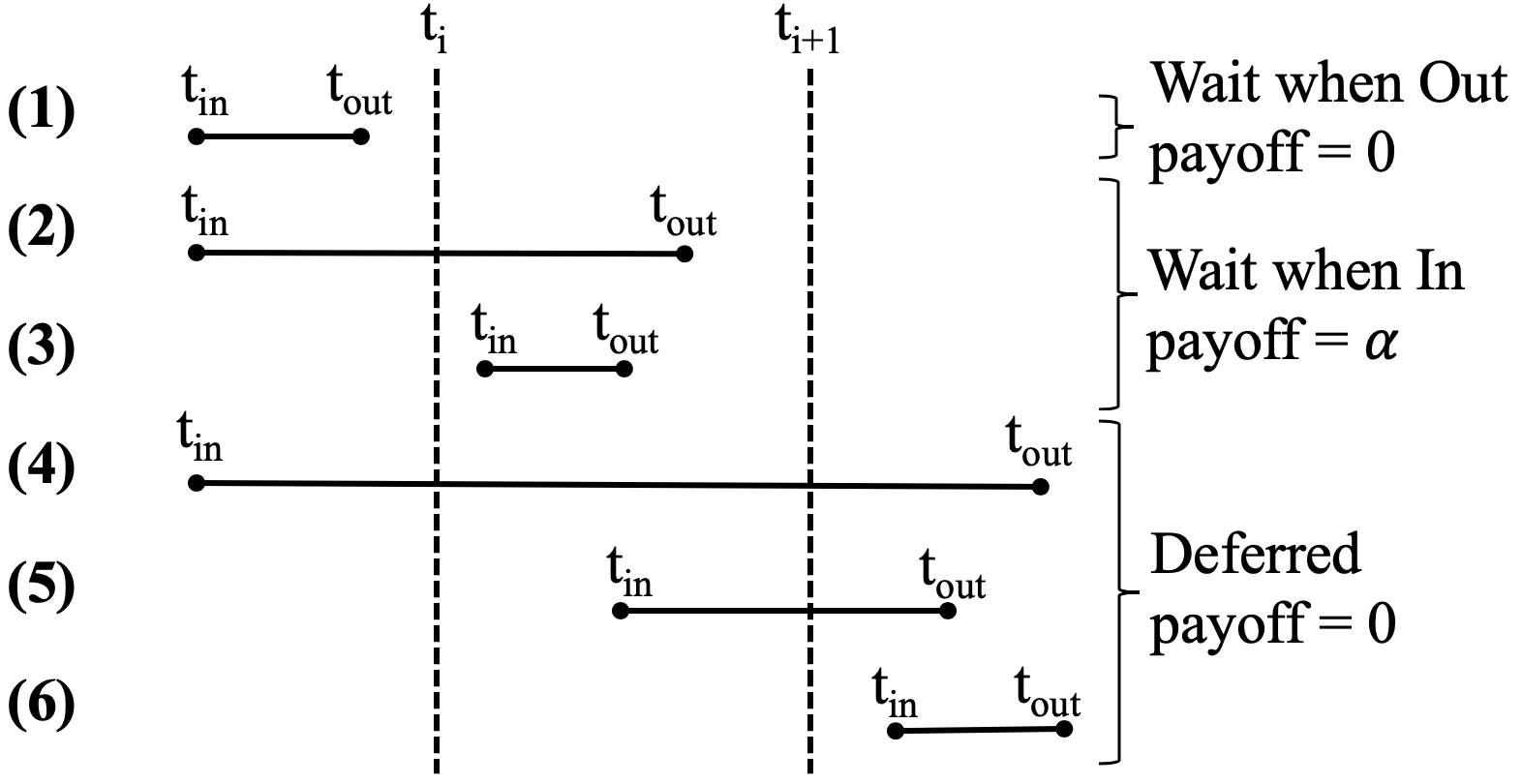}
\caption{Illustrations of cases when we "wait" (\emph{i.e.}, $t_{i+1} < \hat{t}$).\label{fig:alpha_cases.PNG}}
\end{figure}

In the case of $t_{i+1} < \hat{t}$, which means we "wait" before the next measurement arrives, we further investigate different cases of $[t_{in}, t_{out}]$ and $[t_{i}, t_{i+1}]$ to decide the reward/penalty. These cases are illustrated in Figure~\ref{fig:alpha_cases.PNG}.
In case (1) where $t_{out} < t_{i}$, we "wait" while the user is outside, thus, the payoff is $0$.
In cases (2) and (3) where $t_i \leq t_{out} \leq t_{i+1}$, we "wait" while the user is inside, thus the payoff is $\alpha$.
In cases (4), (5), and (6) where have $t_{i+1} < t_{out}$, although the user might be inside, we still have a chance to act when the next measurement arrives. Therefore, the evaluation is deferred to the next measurements and the payoff of the current measurement is $0$. In short, we only receive payoff $\alpha$ when $t_i \leq t_{out} \leq t_{i+1}$. This explains the $\alpha$ payoff in Equation~\ref{eq:realized_value_as_a_function_of_t}.

Recall that $\hat{t}$ in Equation~\ref{eq:realized_value_as_a_function_of_t} comes from the location prediction in Section~\ref{subsec:location_prediction} and expected value computations from Equation~\ref{eq:expected_payoff} in Section~\ref{subsec:decision_theory}. The computed value of $\hat{t}$ varies depending on various algorithmic choices, such as the kernel function of the GP. Different values of $\hat{t}$ affect $V_{\mathcal{R},\mathcal{S}}(t_i)$, so we can assess the effect of our algorithmic choices.

For a given trajectory, we should only act on a geofence a maximum of one time. Thus, for evaluation purposes, we stop calculating the realized value for this trajectory after the measurement $\bm{x}(t_k)$ for which we receive $\beta$ (\emph{i.e.} act when inside the geofence), and set $V_{\mathcal{R},\mathcal{S}}(t_i)$ of all following measurements to $0$, \emph{i.e.} $ \forall i \in [k+1, N_{\mathcal{S}}-1]: V_{\mathcal{R},\mathcal{S}}(t_i) = 0$. This means we can receive the payoff $\beta$ only once.

The realized value of a whole trajectory $\mathcal{S}$ is then the sum of realized values of all the measurements of $\mathcal{S}$:
\begin{equation}
    V_{\mathcal{R},\mathcal{S}} = \sum_{i = 1}^{N_{\mathcal{S}} - 1} V_{\mathcal{R},\mathcal{S}}(t_i)
\end{equation}

To avoid any bias towards any geofence positioning, we consider sum of $V_{\mathcal{R},\mathcal{S}}$ of a trajectory $\mathcal{S}$ over a set of multiple geofences as the score of $\mathcal{S}$:
\begin{equation}
    V_{\mathcal{S}} = \sum_{\mathcal{R}} V_{\mathcal{R},\mathcal{S}}
\end{equation}

\noindent Our set of experimental geofences is a grid described below.

The final realized value is the average of $V_{\mathcal{S}}$ over all trajectories:
\begin{equation}
    V = \frac{1}{N_T} \sum_{\mathcal{S}} V_{\mathcal{S}}
\end{equation}
\noindent where $N_T$ is the number of trajectories. The realized value $V$ is a simple way to compare the aggregated payoff of different algorithms, tested over multiple trajectories and geofences.

\subsection{Baselines}
\label{subsec:baseline}
Our baseline algorithm is called Passive Wait (PW), where the geofence passively waits to act until there is an actual location measurement inside. This appears to be the basis of the algorithm used by both iOS~\cite{iOSGeofencing} and Android~\cite{AndroidGeofencing}. Formally, when receiving the most recent measurement $\bm{x}(t_0) = \big(x(t_0), y(t_0)\big)$ at time $t = t_0 = 0$ with the measurement noise defined by the standard deviation $\sigma_m$, then for each $t < t^*$, the PW method makes a prediction:

\begin{equation}
    \boldsymbol{x}(t) \sim \mathcal{N} \big(\boldsymbol{\mu}(t), \boldsymbol{\Sigma} (t)\big) 
            = \mathcal{N}\Bigg(
                \Bigg[\begin{matrix}
                x(t_0)\\
                y(t_0)\\
                \end{matrix}\Bigg], 
                \sigma^2_m I
                \Bigg)
\end{equation}

\noindent where $I$ is the $2 \times 2$ identity matrix. While we call this a prediction, it is simply asserting that the user stays at the measurement point $\boldsymbol{x}(t) = \big[x(t_0),y(t_0)\big]^T$ until the next measurement and that the uncertainty in location is solely due to measurement noise. Although this is a simple baseline algorithm, it appears to be the most sophisticated existing algorithm for geofences, both in practice and in research. We also explore and compare different instantiations of the GP.

\subsection{Experiment Setup}
\label{subsec:exp_setup}
We present several experiments on 530 trajectories from 165 unique users, processed from high-density data described in Section~\ref{sec:experimental-data}. It is difficult to accurately predict the parameters of an actual geofence application, so we experiment with a thorough range of experimental parameters to show our algorithm performs well in a variety of operating conditions.

For each trajectory, we created a grid of geofences centered at the last training point. The grid covered the entire bounding box of the testing data of the trajectory. In order to account for predictions beyond the bounds of the trajectory, we expanded the grid on all four sides by 18 kilometers. This expansion assumed a maximum driving speed of 30m/s (or 108km/h) and a duration of 600 seconds. The choice of 600(s) is explained in our discussion of $\epsilon$ in Section~\ref{subsec:exp_epsilon}.

Each grid cell is considered as a geofence with size $L \times L$. We tested with sizes $L \in \{500, \textbf{1000}, 1500, 2000, 2500\}$ meters for square cells, creating from 230 to 13,824 geofences per grid for evaluation depending on $L$ and the specific trajectory. The bold value indicates the default value used in the experiments where this parameter is fixed. 
The grids have sizes ranging from 1444 $km^2$ to 3456 $km^2$. 
We chose these grid sizes to represent a range of possible scenarios. Our algorithm tends to work better for smaller cells, hence for fairness we did not experiment with sizes below 500 meters on a side. With up to 13,824 geofences, combined with the 530 trajectories, this gives a test size of over 7 million trajectory-geofence pairs.

Each trajectory is sub-sampled with $\lambda \in \{0.25, \textbf{0.5}, 1, 2, 4, 8\}$ with a Poisson time interval $\Delta T$ of one minute. Roughly speaking, these values of $\lambda$ mean we expect to receive one measurement from every $4$ minutes ($\lambda = \frac{1}{4}$) to every $7.5$ seconds ($\lambda = 8$).

Then, for each test measurement, the prediction model is allowed to use all measurements in the last $\omega$ seconds as the training data. $\omega$ is called the maximum look-back range and takes values in $\{200, \textbf{300}, 400, 500, 600\}$ seconds. 
The maximum prediction threshold $\epsilon$, which is used to find the maximum prediction time $t^*$, is set to $\epsilon \in \{0.1, \textbf{0.2}, 0.3, \dots, 0.9\}$.

Two settings for the payoff matrix is considered to represent two real-world use cases: the \textit{advertising} matrix where $\alpha, \beta, \delta = -\frac{1}{2}, 1, -\frac{1}{4}$, respectively; and the \textit{alert-zone} matrix where $\alpha, \beta, \delta = -2, 1, -\frac{1}{4} $, respectively. For the advertising case, the geofence messenger wants to deliver an advertisement to users inside the geofence. The alert-zone case involves a messenger who wants to deliver an important message, e.g., a safety warning.
The intuition behind those values of $\alpha, \beta, \delta$ is that the value received when we act correctly ($\beta$) is often higher than the penalty when we act incorrectly (\emph{i.e.}, $|\beta| > |\delta|$); and when the user is inside a geofence, there are much more hazardous consequences if we fail to act in case of alert zone than advertising (\emph{i.e.}, $|\alpha_{\textit{alert-zone}}| > |\alpha_{\textit{advertising}}|$). While these values were set based on our own reasoning to mimic advertising and alert-zone scenarios, we assume the payoff matrix is given or can be learned. We instead focus on the problem of making decisions for geofence activations.

We use the squared exponential kernel from Section~\ref{subsec:location_prediction} as the default kernel due to its popularity. Other popular kernels, such as Rational Quadratic and Matérn kernels\cite{Rasmussen2005GPM}, also achieve comparable performance but we omit those results due to space limitations.

\subsection{Experimental Results}

\subsubsection{Varying Poisson $\lambda$}

\begin{figure}[htbp!]
\centering
\begin{subfigure}{.5\linewidth}
  \centering
  \includegraphics[width=\linewidth]{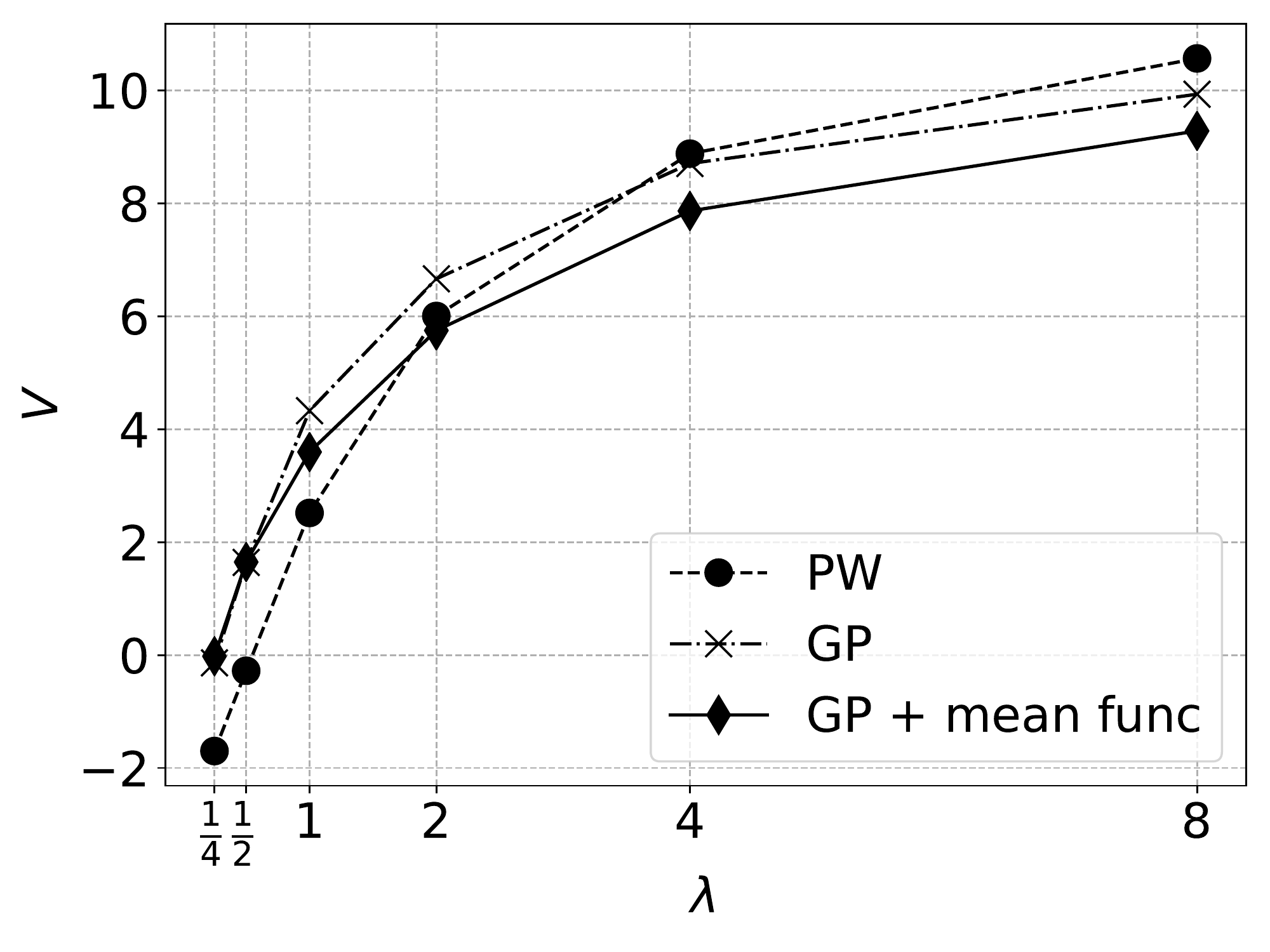}
  \caption{Advertising}
  \label{fig:vary_lambda_advertising}
\end{subfigure}%
\begin{subfigure}{.5\linewidth}
  \centering
  \includegraphics[width=\linewidth]{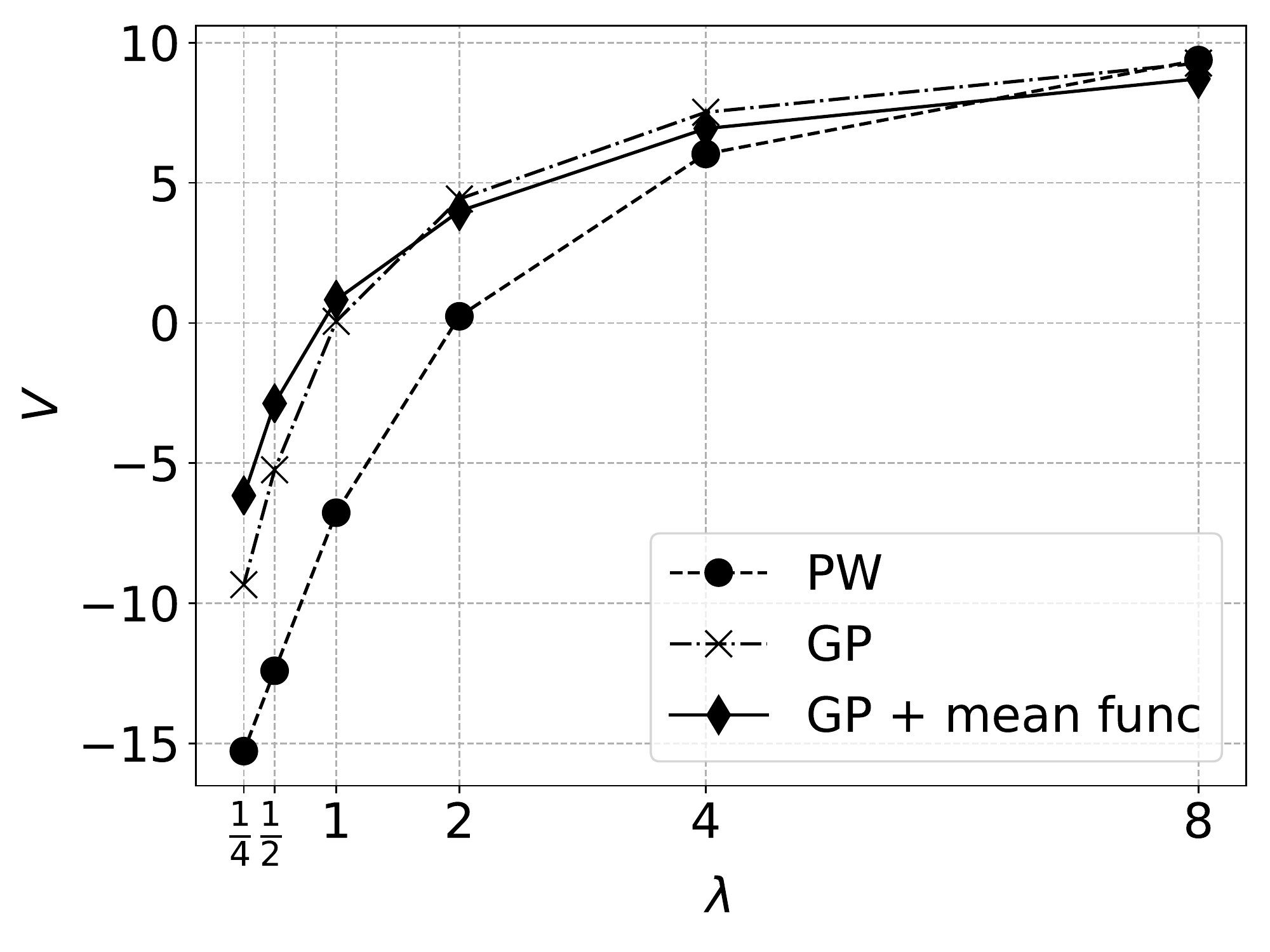}
  \caption{Alert-zone}
  \label{fig:vary_lambda_alertzone}
\end{subfigure}
\caption{The realized value with different values of $\lambda$}
\label{fig:vary_lambda}
\end{figure}

As we focus on the sporadic setting, we first evaluate the methods for different values of $\lambda$. Figure~\ref{fig:vary_lambda} shows the realized values $V$ when $\lambda$ varies and other parameters are fixed as the bold values shown in Section~\ref{subsec:exp_setup}, for both advertising and alert-zone payoff matrices. Roughly speaking, each data point in a result graph shows the average dollar amount per trajectory that one model achieved over all geofences. As a concrete example, in Figure~\ref{fig:vary_lambda_advertising}, the realized value of GP + mean func for $\lambda = 4$ is about 8. This means that GP + mean func achieved \$8 benefit per trajectory summed over all geofences.

The general observation is that GP-based algorithms greatly outperform PW, especially with smaller $\lambda$, \emph{i.e.}, more sporadic. 
This confirms our hypothesis that for sporadic measurements, using principled decision theory with a proper location prediction method brings significant improvement.

The improvement is even more prominent in the case of the alert-zone payoff matrix in Figure~\ref{fig:vary_lambda_alertzone},
where the penalty is higher if the messenger waits until the user is inside the dangerous area (\emph{i.e.}, larger $|\alpha|$). The reason is that we tend to act more readily when $\alpha$ is larger, while PW does not. We can gain more insight into this improvement by investigating our decision making process further. We decide to act when:
\begin{equation}
    \mathbb{E}\big[V(t) \mid \textrm{act}\big] > \mathbb{E}\big[V(t) \mid \textrm{wait}\big]
\end{equation}
\noindent which means:
\begin{equation}
    \beta p_\mathcal{R}(t) + \delta (1-p_\mathcal{R}(t)) > \alpha p_\mathcal{R}(t) + 0(1-p_\mathcal{R}(t)) 
\end{equation}
\noindent After re-arranging with the fact that typically $\delta + \alpha - \beta < 0$:
\begin{equation}
    p_\mathcal{R}(t) > \frac{\delta}{\delta + \alpha - \beta}
    \label{eq:condition_act}
\end{equation}
\noindent When $|\alpha|$ increases with fixed $\delta$ and $\beta$, the value of the right hand side of Equation~\ref{eq:condition_act} becomes smaller, which means that we might act with a smaller probability $p_\mathcal{R}(t)$. In our specific experimental settings, this probability is $\frac{1}{7}$ for advertising and $\frac{1}{13}$ for alert-zone. Therefore, an advertising act needs a larger probability of being inside the geofence than an alert-zone act, because missing an alert-zone act is is more costly.

When location measurements are frequent (\emph{i.e.}, larger $\lambda$), prediction does not offer much benefit. This is understandable because with frequent measurements, there is a much higher chance a measurement arrives when the user is inside the geofence. Therefore, the prediction might be unnecessary.

\begin{figure}[htbp!]
\centering
\begin{subfigure}{.5\linewidth}
  \centering
  \includegraphics[width=\textwidth]{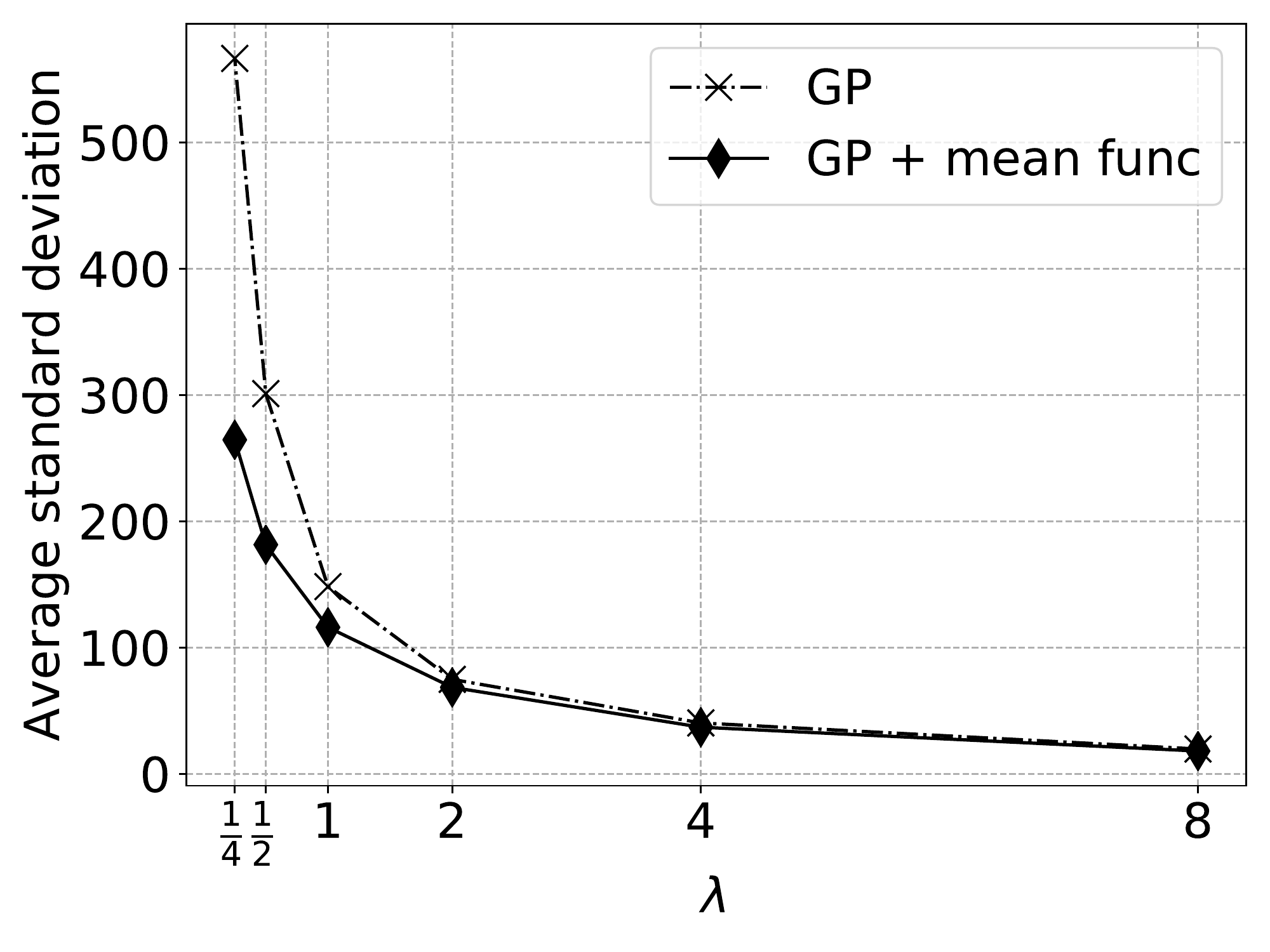}
  \caption{Average standard deviation per prediction}
  \label{fig:vary_lambda_avg_std}
\end{subfigure}%
\begin{subfigure}{.5\linewidth}
  \centering
  \includegraphics[width=\textwidth]{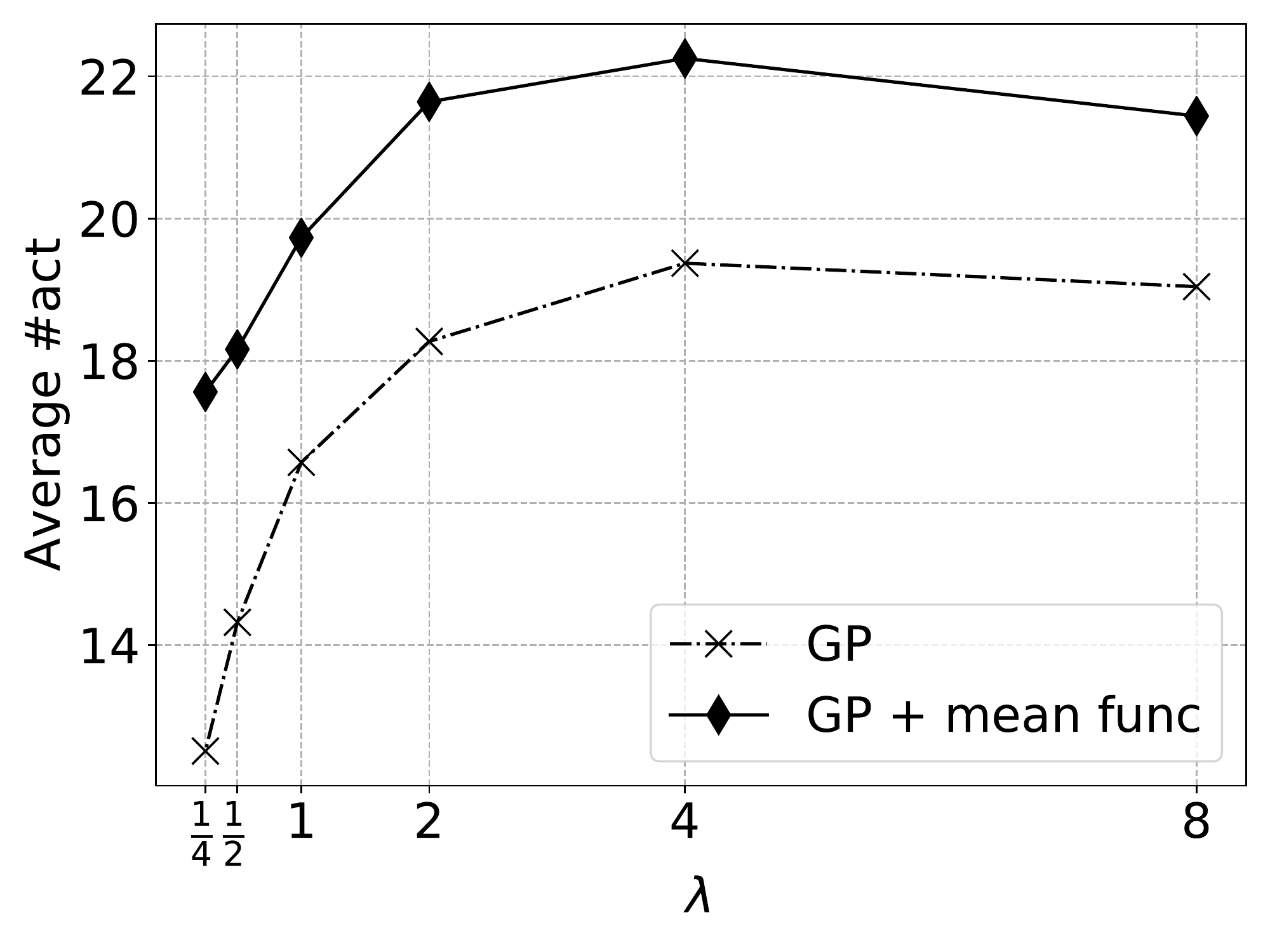}
  \caption{Average number of acts per trajectory (alert-zone setting)}
  \label{fig:vary_lambda_avg_atc}
\end{subfigure}
\caption{The comparison between the standard GP and the GP with linear mean function}
\label{fig:vary_lambda_avg_std_act}
\end{figure}

The modified version of the GPs with a linear mean function, denoted as "GP + mean func" in the figures, offers a better realized value in the alert-zone setting compared to the standard GP. This improvement is also shown in other following experiments.
Besides the intuition that the movement in the near future tends to rely more on the recent locations, another potential explanation is that the modified GP produces predictions with smaller variance, because the line going through the most recent measurements might be a better mean function than a zero mean. The smaller variances, along with the aforementioned analysis of $p_\mathcal{R}(t)$, might give a higher number of acts. Therefore, in the alert-zone setting where the risk is higher for waiting than acting, the modified GP can achieve a higher payoff. 
Figure~\ref{fig:vary_lambda_avg_std_act} shows the comparison between the standard GP and the modified GP. 
First, Figure~\ref{fig:vary_lambda_avg_std} shows the average value of the standard deviation per prediction. This value of the modified GP is smaller than that of the standard GP, and they become closer to each other when $\lambda$ increases. Second, in Figure~\ref{fig:vary_lambda_avg_atc} we can see a clear gap between the number of "act" decisions of the modified GP and that of the standard GP in the alert-zone setting. However, again, these differences do not play an important role for the performance when the measurements arrive frequently (\emph{i.e.}, large $\lambda$).

\subsubsection{Varying Prediction Threshold $\epsilon$}
\label{subsec:exp_epsilon}
Next, we consider the effect of the maximum prediction threshold $\epsilon$. Recall that $\epsilon$ is combined with $\lambda$ to decide the maximum prediction time $t^*$ as the principled way of taking into account the fact that the next location measurement may come at any time in the future, which would suspend the current prediction. For a fixed $\lambda$, a smaller $\epsilon$ leads to a larger $t^*$. 

\begin{figure}[htbp!]
\centering
\includegraphics[width=.5\linewidth]{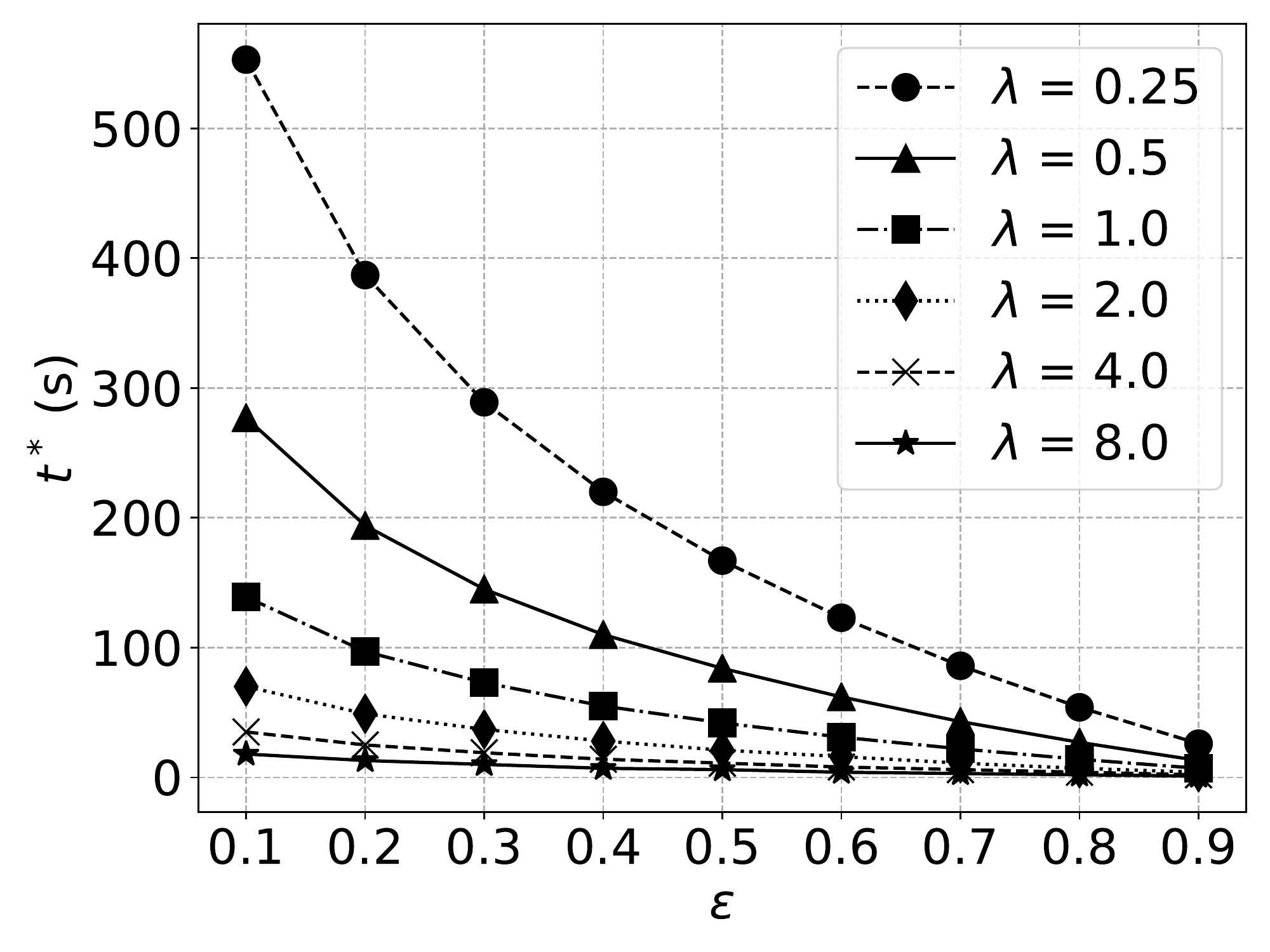}
\caption{Maximum prediction time $t^*$ given $\epsilon$.\label{fig:max_prediction_range_from_lambda}}
\end{figure}

The corresponding $t^*$ for each value of $\epsilon$ is shown in Figure~\ref{fig:max_prediction_range_from_lambda}.
The general observation is that we need to make predictions further ahead in time (\emph{i.e.}, larger $t^*$) when data is more sporadic and/or $\epsilon$ is smaller, and vice versa.
In our most sporadic setting, where $\lambda = \frac{1}{4}$, and most conservative threshold, $\epsilon = 0.1$, we need to make predictions up to $600s$ ahead. That explains why $600s$ is used in our expansion of the geofence grid discussed in Section~\ref{subsec:exp_setup}.

\begin{figure}[htbp!]
\centering
\begin{subfigure}{.5\linewidth}
  \centering
  \includegraphics[width=\textwidth]{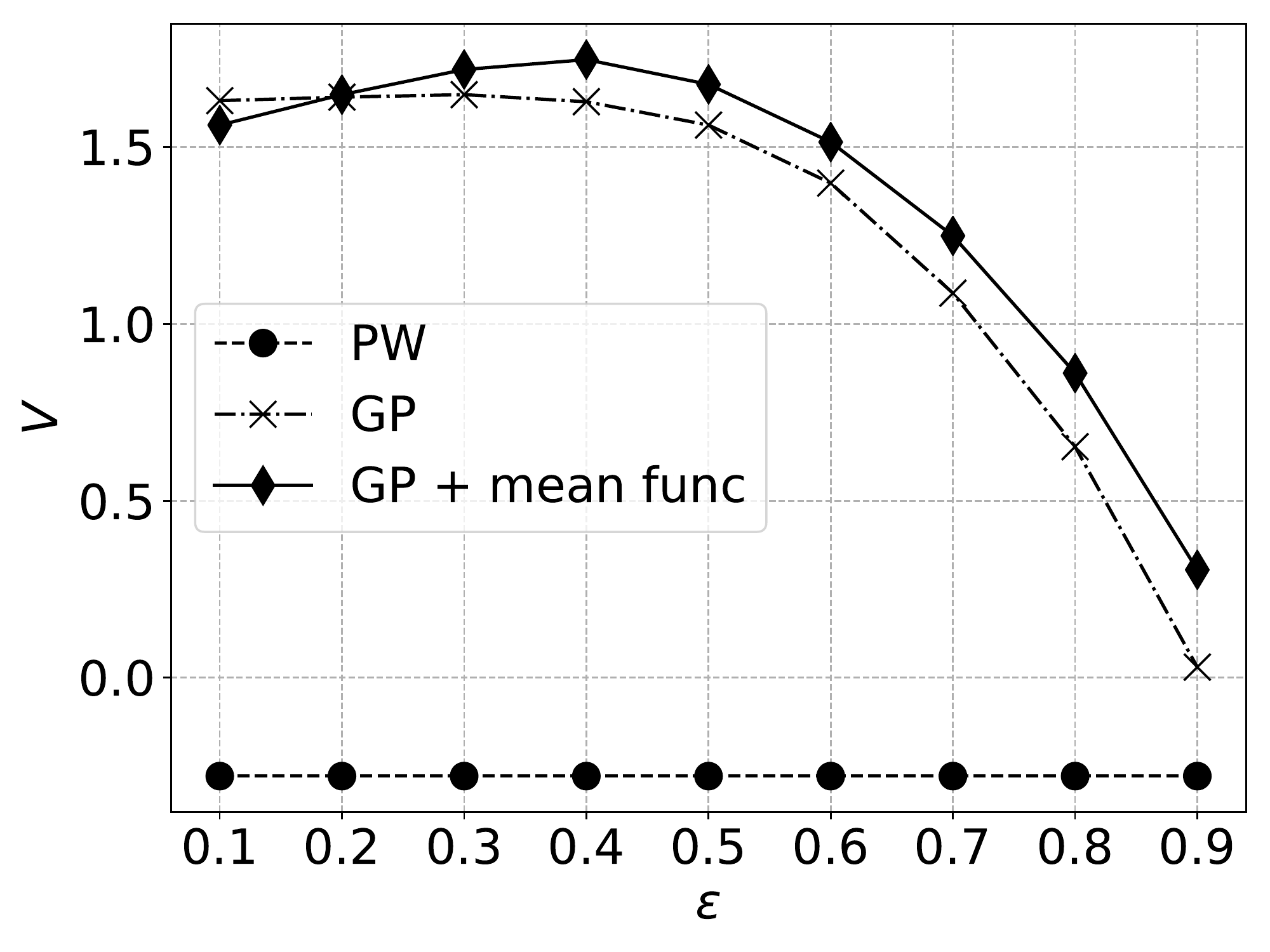}
  \caption{Advertising}
  \label{fig:vary_max_pred_threshold_eps_advertising}
\end{subfigure}%
\begin{subfigure}{.5\linewidth}
  \centering
  \includegraphics[width=\textwidth]{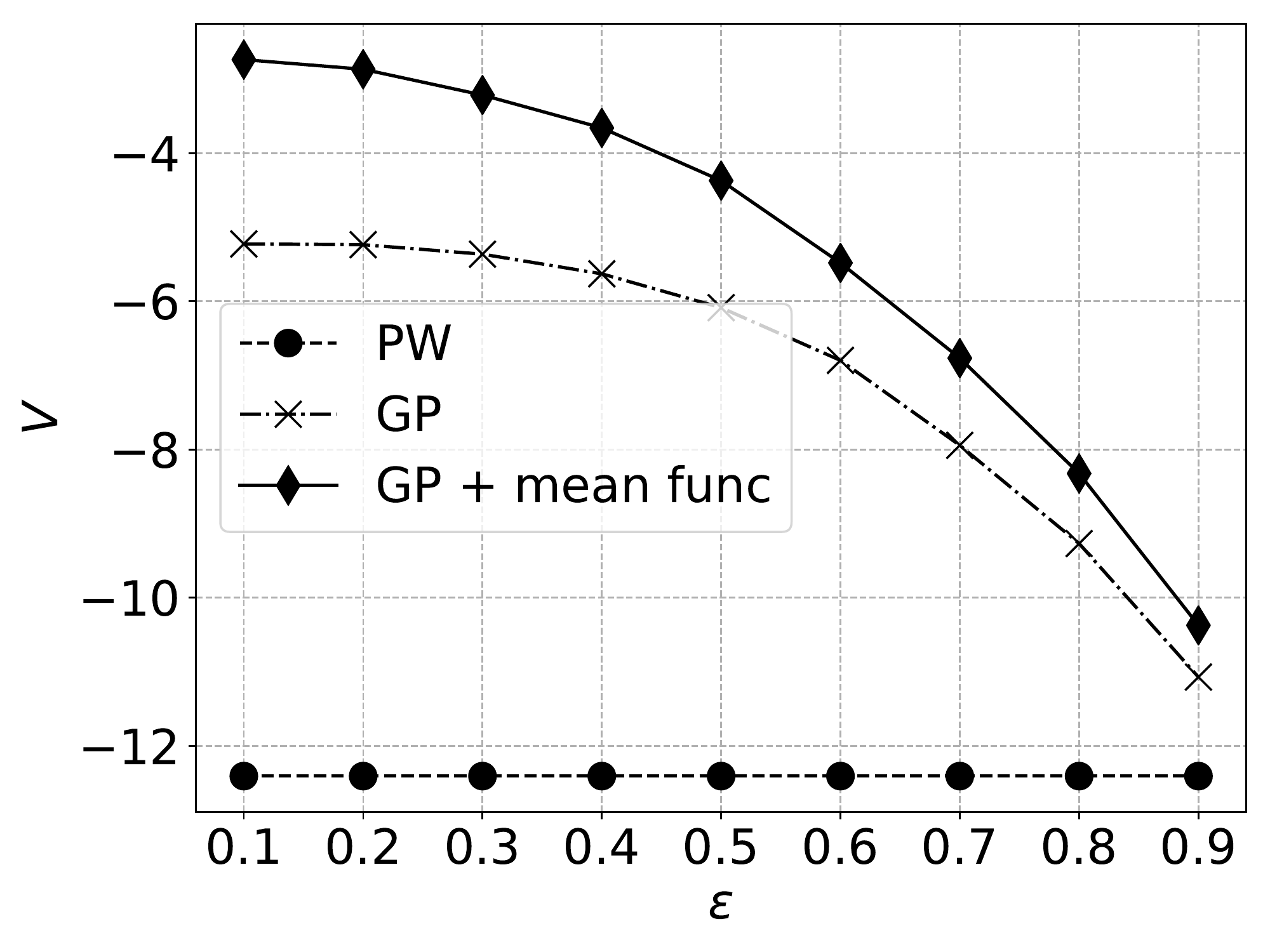}
  \caption{Alert-zone}
  \label{fig:vary_max_pred_threshold_eps_alertzone}
\end{subfigure}
\caption{The realized values when varying $\epsilon$}
\label{fig:vary_max_pred_threshold_eps}
\end{figure}

Figure~\ref{fig:vary_max_pred_threshold_eps} shows the realized values for different values of $\epsilon$ when $\lambda$ is fixed at $\frac{1}{2}$. 
The general trend is that smaller values of $\epsilon$ give larger realized values. With larger $\epsilon$, the realized value becomes smaller and closer to the value obtained by the PW, which does not make predictions, thus, is not affected by the change of $\epsilon$. In general, using a smaller $\epsilon$ is more conservative, because it leads to longer prediction times, lessening the chance that we will terminate the location prediction before the next measurement arrives.

\begin{table}[htbp!]
\centering
\begin{tabular}{|l||c|c|c|c|c|}
\hline
$t^*$ (s) & 194  & 277 & 600 & 3600 & 36,000  \\ \hline
Time (s)  & 0.38 & 0.53 & 1.08 & 6.12 & 60.52  \\ \hline
\end{tabular}
\caption{Time to calculate $p_\mathcal{R}(t)$ for all $0 < t < t^*$}
\label{tbl:comp_time_epsilon}
\end{table}

Another important effect of $\epsilon$ is that the suspension of the prediction greatly reduces computation time.
In our experiments, running with a single thread on a personal computer 
, the training and prediction time of the GPs are about $1.7$s and $0.6$s, respectively. These times are independent of $\epsilon$.
However, for prediction, for all $0 < t < t^*$, we need to calculate the integrals on the predicted location distributions to calculate  $p_\mathcal{R}(t)$ as in Equation~\ref{eq:probability_inside_geofence}.
Table~\ref{tbl:comp_time_epsilon} shows the time required to calculate $p_\mathcal{R}(t)$ for all $0 < t < t^*$, for different values of $t^*$ with $\lambda = \frac{1}{2}$, and for a single geofence. While all these values of $t^*$ result in similar realized values, with $\epsilon = 0.1$ and $0.2$, we can suspend the prediction at $t^* = 194$s and $277$s, and the calculation time is just $0.38$s and $0.53$s, respectively. Any computation after these values of $t^*$ can be considered as wasted computation.
Without this principled approach to identify $t^*$, one could arbitrarily choose some values for $t^*$, which would result in an almost linear increase in the computation time. For $t^* = 3600$s and $36,000$s, the computation times are $6$s and $60$s, respectively, which are often not suitable for most real-time mobile applications.

\subsubsection{Varying Geofence Size $L$}

\begin{figure}[htbp!]
\centering
\begin{subfigure}{.5\linewidth}
  \centering
  \includegraphics[width=\textwidth]{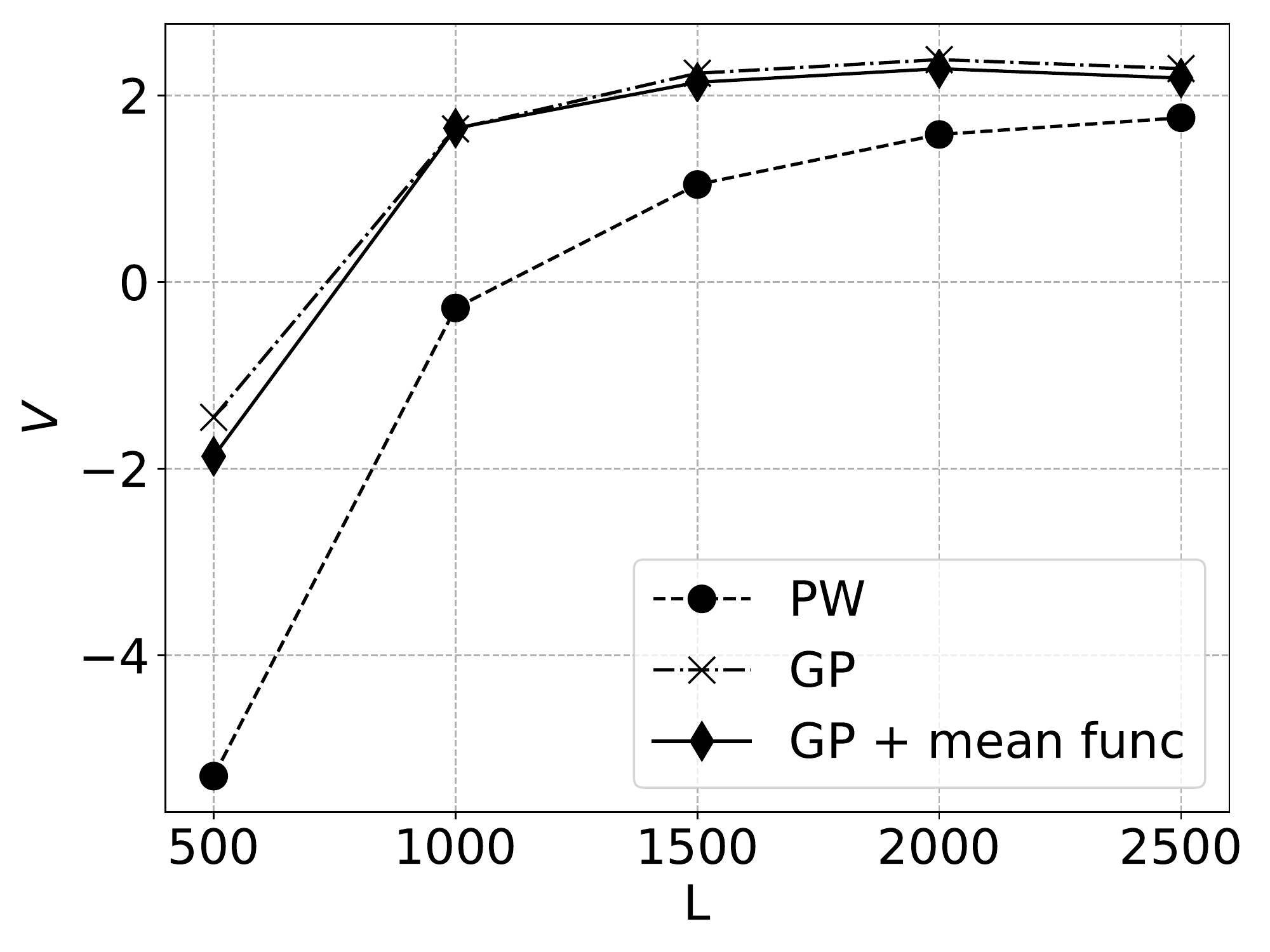}
  \caption{Advertising}
  \label{fig:vary_geofence_size_advertising}
\end{subfigure}%
\begin{subfigure}{.5\linewidth}
  \centering
  \includegraphics[width=\textwidth]{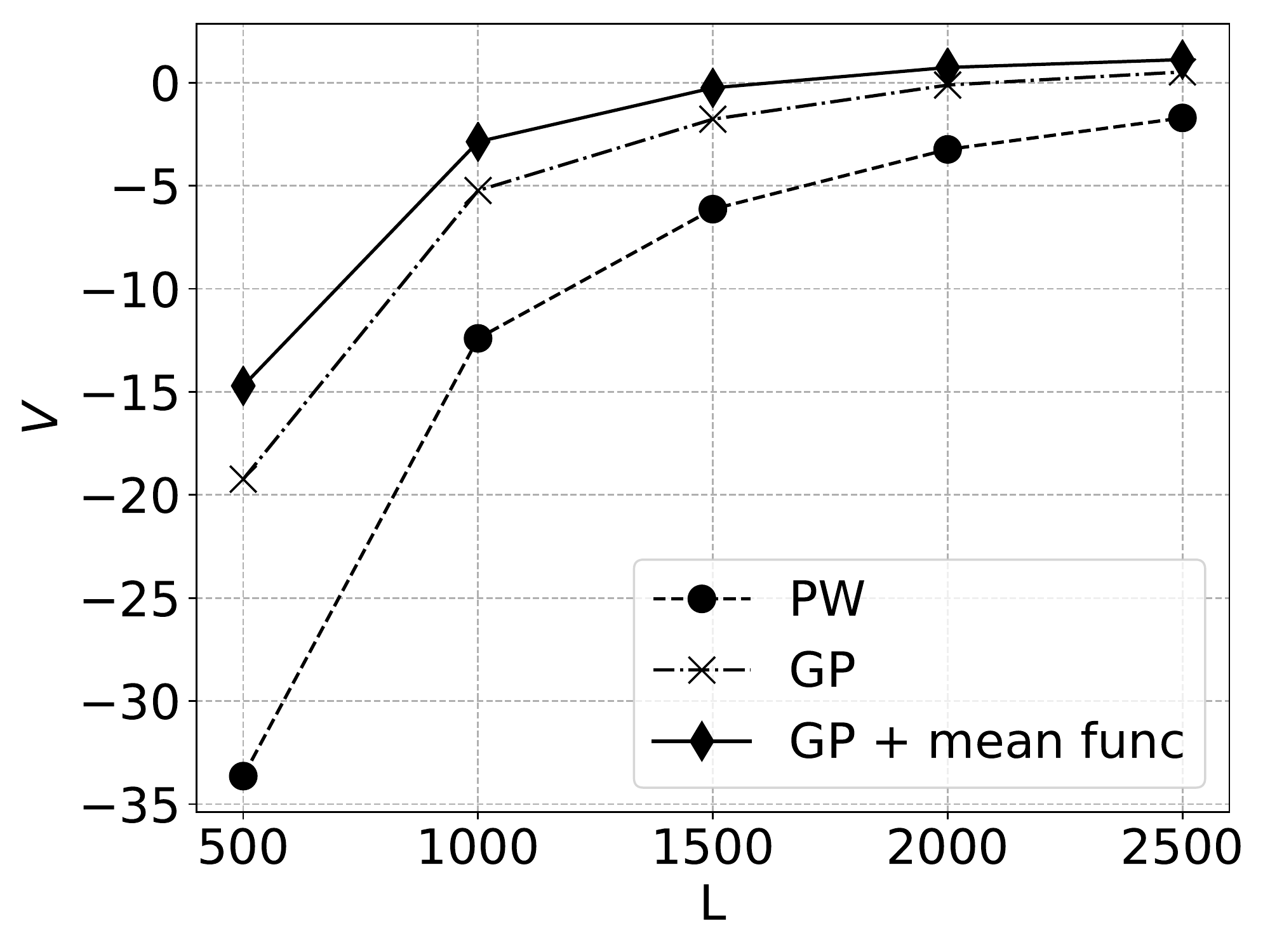}
  \caption{Alert-zone}
  \label{fig:vary_geofence_size_alertzone}
\end{subfigure}
\caption{The realized values when varying geofence size $L$}
\label{fig:vary_geofence_size}
\end{figure}

Next, we evaluate the performance with different sizes of the geofences. The result is shown in Figure~\ref{fig:vary_geofence_size}. We can see a clear advantage of GPs over PW in smaller geofence sizes. The reason is that with smaller geofences, there is a higher chance that we might not receive any measurement when the user is inside. Thus, the PW approach would miss those geofences while the prediction of the GPs can still help us take the right action. On the other hand, when the size of the geofence is larger, then, again, there is a higher chance a measurement arrives inside the geofence even though the measurements do not come frequently. However, the prediction of GPs still gives a better realized value than the PW.

\subsubsection{Varying Look-Back Range $\omega$}
\begin{figure}[htbp!]
\centering
\begin{subfigure}{.5\linewidth}
  \centering
  \includegraphics[width=\textwidth]{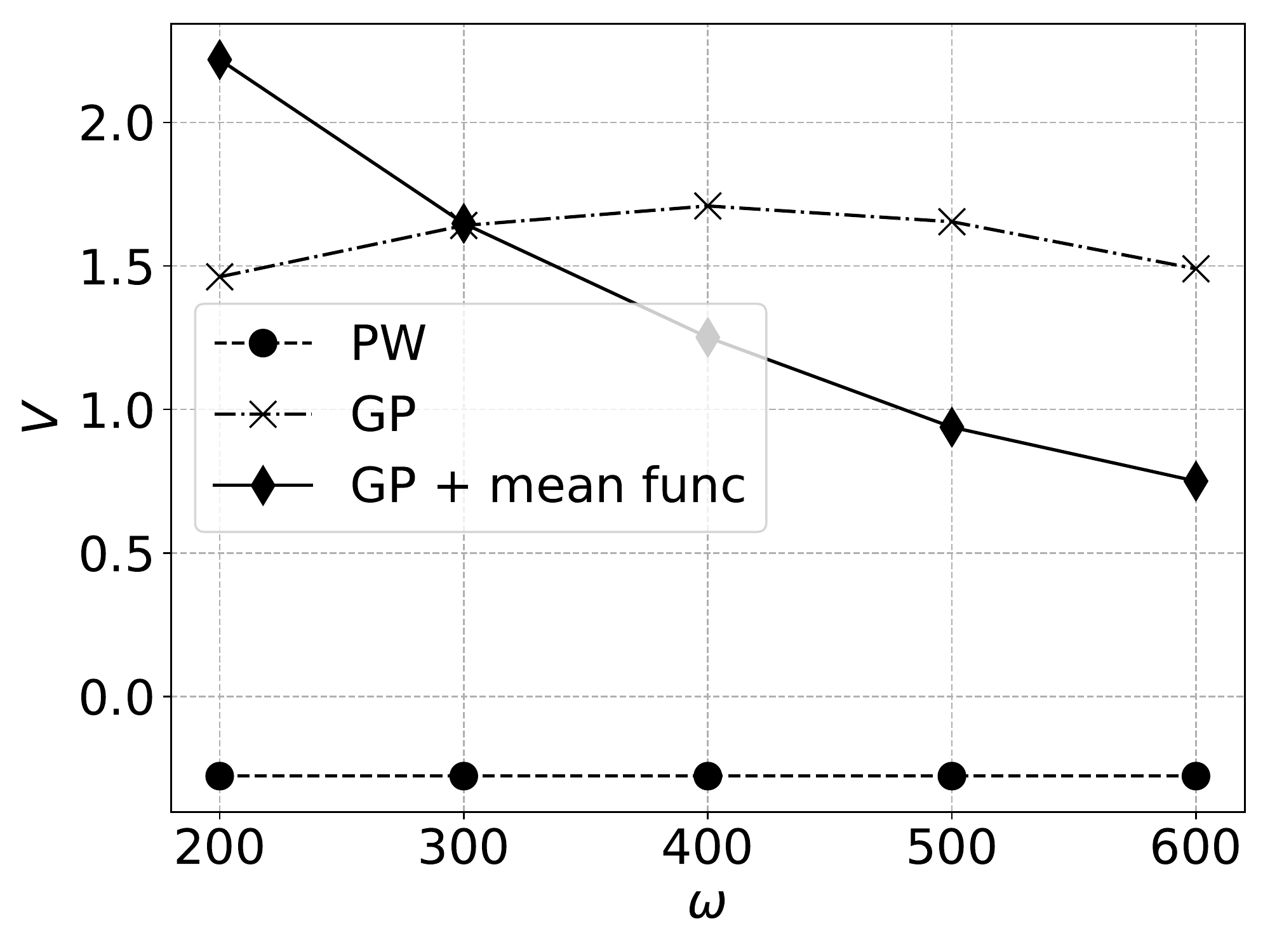}
  \caption{Advertising}
  \label{fig:vary_lookback_range_advertising}
\end{subfigure}%
\begin{subfigure}{.5\linewidth}
  \centering
  \includegraphics[width=\textwidth]{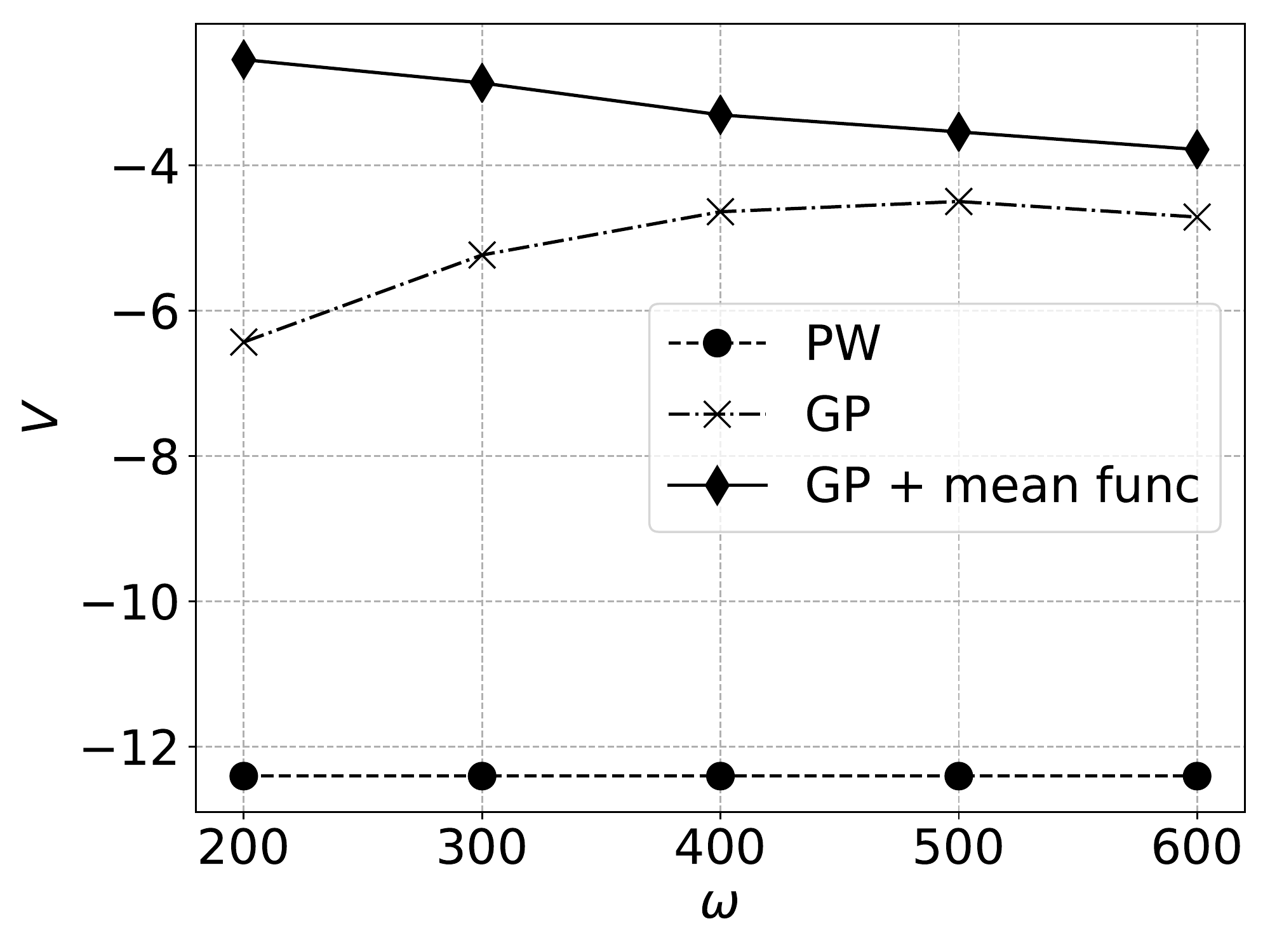}
  \caption{Alert-zone}
  \label{fig:vary_lookback_range_alertzone}
\end{subfigure}
\caption{The realized values when varying $\omega$ (in seconds)}
\label{fig:vary_lookback_range}
\end{figure}

The effect of the maximum look-back range $\omega$ is shown in Figure~\ref{fig:vary_lookback_range}. This parameter controls how far back in time we look for training data to train the parameters of the GP. While both GP-based methods still outperform PW, when $\omega$ becomes too large (\emph{e.g.}, 600s), it negatively affects the performance of the GP-based methods. One possible explanation is that the movement pattern too far into the past may not relate well to the movement in the near future.
Also the change of $\omega$ shows a greater effect on the GP with the linear mean function than the normal GP. This is because the linear mean function further drives the prediction towards the linear extrapolation of the most recent measurements, thus making a larger difference between the pattern in the far past and the near future.

\subsubsection{Varying Measurement Noise $\sigma_m$}

\begin{figure}[htbp!]
\centering
\begin{subfigure}{.5\linewidth}
  \centering
  \includegraphics[width=\textwidth]{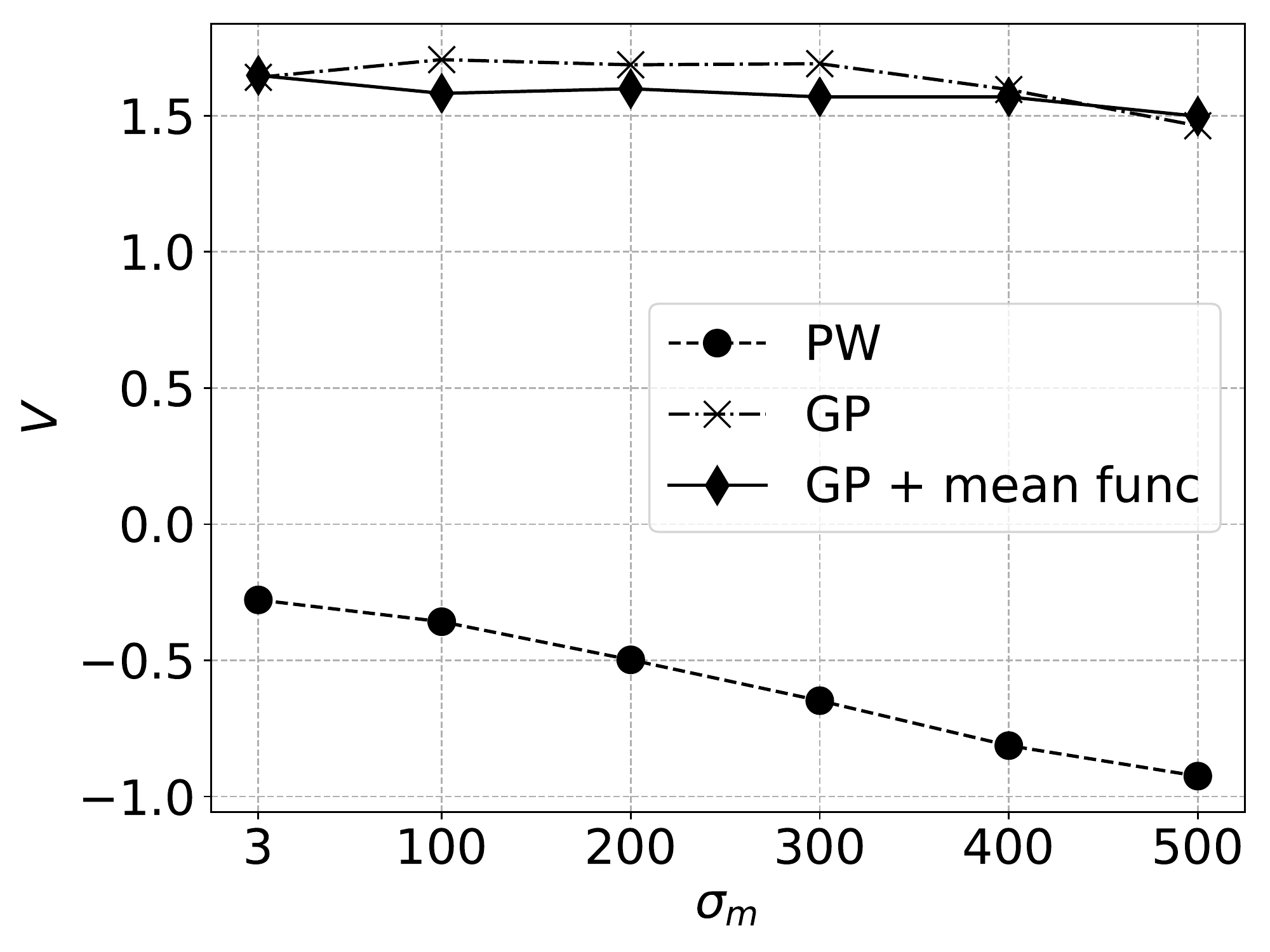}
  \caption{Advertising}
  \label{fig:vary_measurement_std_advertising}
\end{subfigure}%
\begin{subfigure}{.5\linewidth}
  \centering
  \includegraphics[width=\textwidth]{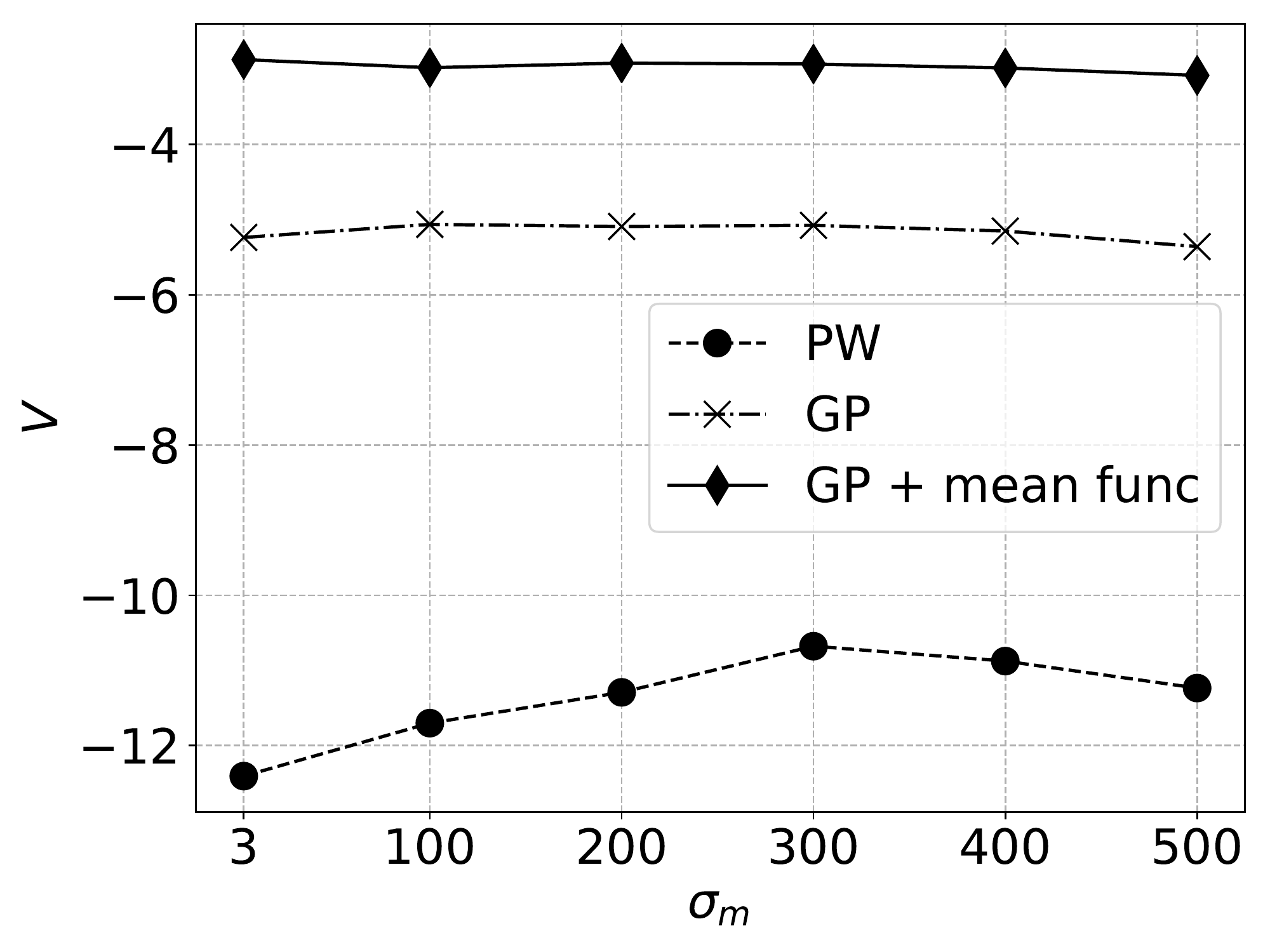}
  \caption{Alert-zone}
  \label{fig:vary_measurement_std_alertzone}
\end{subfigure}
\caption{The realized values when varying measurement standard deviation $\sigma_m$ (in meters)}
\label{fig:vary_measurement_std}
\end{figure}

Next, we investigate the effect of the measurement noise given by the standard deviation $\sigma_m$ in Figure~\ref{fig:vary_measurement_std}. In our algorithm, we must set the value of $\sigma_m$ as an estimate of the measurement noise. In Figure~\ref{fig:vary_measurement_std_advertising} and~\ref{fig:vary_measurement_std_alertzone}, $\sigma_m$ is varied from a reasonable GPS error of $3$m to a very noisy $500$m. These values of $\sigma_m$ are provided to the prediction methods. The GP-based methods successfully capture the pattern and bring superior realized values compared to the standard approach of PW.

\section{Conclusions and Future Work}
While geofences have proven useful for decades, the problem of using them with sparsely sampled, sporadic data has not received attention in the research literature. If location is not sampled frequently enough, it increases the chances of missing a geofence. We proposed a fundamentally new approach to this problem from a probabilistic and utility-based perspective. We presented a conceptually simple solution that predicts a user's location between measurements. The prediction gives a probability distribution over locations that changes with time, and thus the predicted probability of being inside the geofence changes with time. This uncertainty is combined with a payoff matrix that allows us compute the expected value of either triggering the geofence's preprogrammed action or instead waiting and doing nothing. We used Gaussian processes for our location predictions, but any prediction that gives a probability distribution as a function of time would work in our framework. 

The prediction begins after each measurement and continues until the next measurement. Since the arrival time of the next measurement is unknown, we presented a method to stop the prediction based on Poisson-distributed arrival time for the next measurement.
Our technique leads to a light-weight model that can operate on a mobile device, without triggering new measurements nor transmitting measurements out of the device. To the best of our knowledge, none of our major contributions (which includes decision theory, probabilistic location prediction, and reasoning about sporadic measurements) has appeared in the research literature of geofences before.

While our algorithm consistently outperforms the baseline over a variety of settings, there are still opportunities for further work in this area. Example directions include investigating other techniques for short term location prediction such as Kalman filters and techniques exploiting the road network; moving geofences; or a better method to decide when to trigger, such as such as when entry into a geofence reaches a certain confidence level.

\bibliographystyle{IEEEtran}
\bibliography{bibliography}

\end{document}